# SUPERNOVA BLASTWAVES AND PRE-SUPERNOVA WINDS: THEIR COSMIC RAY CONTRIBUTION

Peter L. Biermann

*Max Planck Institut für Radioastronomie, Bonn, Germany*

Shocks in stellar winds can accelerate particles; energetic particles are observed through nonthermal radioemission in novae, OB stars and Wolf Rayet stars. Supernova explosions into predecessor stellar winds can lead to particle acceleration, which we suggest can explain most of the observed cosmic rays of the nuclei of Helium and heavier elements, from GeV in particle energies up to near $3\,10^9$ GeV, as well as electrons above about 30 GeV. We go through the following steps to make the case: 1) Using a postulate for an underlying principle that leads to transport coefficients in a turbulent plasma, we derive the properties of energetic particles accelerated in spherical shocks in a stellar wind. 2) We suggest that a dynamo working in the inner convection zone of an upper main sequence star can lead to high magnetic field strengths, which may become directly observable in massive white dwarfs, massive red giant stars and Wolf Rayet stars. 3) Such magnetic fields may put additional momentum into stellar winds from the pressure gradient of the toroidal field, with reduced angular momentum loss. 4) We use the statistics of Wolf Rayet stars and radiosupernovae to derive a lower limit for the magnetic field strengths. This limit gives support to the wind driving argument as well as the derivation of the maximum particle energy that can be reached. 5) From a comparison of the radioluminosities of various stars, radio supernovae, as well as supernova remnants, there appears to be a critical Alfvénic Machnumber for electron injection. With this concept we propose an explanation for the observed proton/electron ratio in galactic cosmic rays at GeV energies. 6) We check the model prediction quantitatively on cosmic ray spectrum and chemical composition against airshower data from a) Akeno, b) a world data set, c) Fly's Eye, and d) against further cosmic ray data available from other experiments. 7) Finally, we summarize various important caveats, and outline important next steps as well as checks as regards the implications of these concepts for stars and stellar evolution.

## I. Introduction

Most supernovae are explosions of massive stars (see, *e.g.* Wheeler 1989), often stars that have a stellar wind prior to the explosion. Thus the physics of these stellar winds becomes important for a discussion of what happens when the star explodes and a shockwave travels down such a stellar wind. Optical and X-ray data have been interpreted as due to shock structures and shockheating (*e.g.* Owocki 1992). Obser-vationally, there is also evidence for shockwaves in stellar winds prior to the explosion, such as suggested by nonthermal radioemission from OB and Wolf Rayet stars, as well by the nonthermal radio emission from the nova GK Per (Seaquist *et al.* 1989). An interpretation of this nonthermal radio emission is an important test for any theory of particle acceleration in shocks in stellar winds.

In this chapter we will review recent work on the acceleration process in shock waves in stellar winds, and argue that a large part of the observed cosmic rays can be attributed to supernova shocks in stellar winds.

## A. Stellar winds

Stellar winds are observed in many cases, low mass stars such as the Sun, as well as high mass stars such as OB stars. For the latter the wind driving can be explained as an effect of radiation pressure (Lucy & Solomon 1970, Castor *et al.* 1975, Pauldrach *et al.* 1986, Owocki 1990). For Wolf Rayet stars the momentum in the wind is generally believed to be too large to be explainable as due to radiation driving in the limit of single scattering, and so, as one possible solution, multiple scattering models have been devised. Magnetic fields have been argued to contribute to the driving through the fast magnetic rotator concept (Cassinelli 1982, 1991); this idea, however, has been criticized on the basis that the corresponding large angular momentum loss would lead to a severe self-limitation of the process (Nerney & Suess 1987) and that therefore the process could not be general. There is a modified magnetic rotator model, for which the Alfvénic surface is close to the stellar surface, and hence the angular momentum loss is strongly reduced (Biermann & Cassinelli 1993). In this case the lifetime is not limited by angular momentum loss, but an initial driving of the wind is required, and the pressure gradient of the tangential magnetic field is argued to provide an amplification of the momentum of the wind.

## B. Cosmic Rays

After the discovery of cosmic rays by Hess (1912) and Kohlhörster (1913), Baade & Zwicky (1934) already proposed that supernova explosions produce cosmic rays. Alfvén (1939) argued early for a local origin in our Galaxy, which is confirmed by the age determinations of the cosmic rays (Garcia-Munoz *et al.* 1977). Fermi (1949, 1954) proposed the basic concept of acceleration still being used. Shklovskii (1953) and Ginzburg (1953) made a convincing case for particle acceleration in supernova remnants. Ginzburg (1953) already emphasized the interesting role of novae; and indeed, the nova GK Per is a test case for the evolution of shocks in winds and their particle acceleration. Hayakawa (1956) proposed that stellar evolution gives rise to an enhancement of heavy elements and pointed out the importance of spallation in the in-terstellar medium; and again, the enrichment found in the cosmic ray contribution from supernova explosions into winds is indeed believed today to be due to just this enrichment. And finally, Cocconi (1956) already argued convincingly that the most energetic cosmic ray particles are from an extragalactic origin; the GRO observations (Sreekumar *et al.* 1993) of a neighboring galaxy, the Small Magellanic Cloud, provided the last and a very strong argument, that indeed the cosmic rays in the lower energy range are not universal, and thus have to be galactic. An early seminal form of some of the ideas expressed in the following can be found in Peters (1959, 1961). A brief historical review is given by Ginzburg (1993). The form of Fermi's process used today was discovered nearly simultaneously by Axford *et al.* (1977), Bell (1978a, b), Blandford & Ostriker (1978), and Krymskii (1977). In this picture the main particle energy gain is from repeated scattering off magnetic irregularities on the two sides of a shock front. Since those two sides can be considered as a continuously contracting system relative to each other, particles that remain in the system gain energy.

There are many important reviews and books on cosmic ray physics: We just mention the classical book by Hayakawa (1969), the new book by Berezinsky *et al.* (1990), and the reviews by Drury (1983), Blandford & Eichler (1987), and Jones & Ellison (1991).

Today there are only some well accepted arguments about the origins of cosmic rays: a) The cosmic rays below about $10^4$ GeV are believed to be predominantly due to the explosion of stars into the normal interstellar medium (Lagage & Cesarsky 1983). b) The cosmic rays from near $10^4$ GeV up to the knee, at $5 \ 10^6$ GeV, are likely predominantly due to explosions of massive stars into their former stellar wind (Völk & Biermann, 1988). c) While this latter claim is not undisputed, there has been certainly no agreement yet on the origin of the cosmic rays of higher particle energy.

Direct observation of cosmic rays, either from gound-based instrumentation, from satellites or from balloons has been the driving input for nearly all considerations in cosmic ray work. These data demonstrate that i) the overall spectrum of cosmic rays is a powerlaw up to an energy, commonly referred to as the knee, near $5 \ 10^6$ GeV, to continue with a steeper powerlaw to the ankle, near $3 \ 10^9$ GeV, with a slight turnup beyond and an apparent cutoff near $10^{11}$ GeV; ii) the spectra are about $E^{-2.74}$ for Hydrogen at moderate energy, and slightly flatter for Helium and heavier elements, while the electron spectrum is consistent with the Hydrogen spectrum at low energy and then changes over to about $E^{-3.3}$; iii) the chemical composition at low energy is crudely similar to that of the interstellar medium, with Hydrogen and Helium underabundant relative to Silicon. All such properties require explanation.

We will assume in the following that the correction from the ob-served cosmic ray spectrum to the source spectrum is a change in spectral slope by exactly 1/3 for relativistic particles (see Biermann 1993a, 1994a, b, c), so that we are looking at source spectra of approximately $E^{-2.4}$ below the knee, and of approximately $E^{-2.8}$ above the knee. This can be argued on the basis of a Kolmogorov spectrum of the irregularities of the interstellar magnetic field, from plasma simulations, from analogies with in situ measurements of the solar wind, and other observations of the interstellar medium (see, *e.g.*, Biermann 1993a). It is clear, however, that a Kolmogorov spectrum of interstellar turbulence is insufficient to explain in a straightforward manner, *e.g.*, the abundance ratios in cosmic rays (see, *e.g.*, Garcia-Munoz *et al.* 1987, and Biermann 1994c). Such properties of cosmic rays may require a deeper understanding of the interstellar medium than we currently have. However, we note that the secondary to primary ratio of spallation products in cosmic rays does not simply yield a spectrum of interstellar turbulence in a medium, which varies on time scales equivalent to those of cosmic ray transport, and which has known inhomogeneities of density contrasts of many powers of ten.

## C. Shocks in winds

There are various kinds of evidence for the presence of shock structures in stellar winds, from in situ observations in the solar wind, to optical line profiles to X-ray emission from massive stars (Lucy 1982, Owocki & Rybicki 1985, Owocki *et al.* 1988, Owocki 1990, 1992, 1994, MacFarlane & Cassinelli 1989, Usov & Melrose 1992, Feldmeier 1993). Furthermore, there are shocks to be expected in the colliding wind zones between the binary star components of two massive stars (Eichler & Usov 1993); we do not consider such cases. Here we wish to concentrate on particle acceleration and the ensuing nonthermal radio emission from shock zones in the winds of single stars.

Nonthermal radioemission is usually attributed to the synchrotron emission of energetic electrons gyrating in magnetic fields. OB and Wolf Rayet stars are both stars with strong powerful winds; they are usually thermal radio emitters from free-free emission; and in many cases, they are also nonthermal radio emitters. Similarly, nonthermal radioemission is detected from some novae, in particular, from the nova GK Per. Thus, any theory to explain particle acceleration in spherical shocks has to account for the properties of such nonthermal radio emission; if, as a result, one finds an explanation for a component of cosmic rays, it provides support for the coherence of a theory to account for the origin of cosmic rays.

For this nonthermal radio emission, we can determine its luminosity, spectrum, time dependence, spatial distribution, and, what turns out to be especially useful, its polarization. In supernova remnants and in the nova GK Per we can even determine the spatial arrangement of polarization: This allows us to derive the topology of the magnetic field. For young supernova remnants this spatial arrangement is always predominantly radial, while the compression of an arbitrarily arranged interstellar magnetic field would naturally lead to a tangential configuration. Also, for the wind of the nova GK Per, we expect an unperturbed tangential magnetic field, and a strengthening of such a field in a shock wave; the observations also in this case show a radial field (Seaquist *et al.* 1989). The observed topology of the magnetic field is thus at right angles to that expected on very simple grounds. This implies a very general reason, and we would would like to suggest that this is best understood as the consequence of a rapid convective radial motion in the shocked region of the plasma. If this notion is the correct interpretation of these data, then the diffusive transport of energetic particles in the shocked region is possibly dominated by this rapid convective motion, and its properties then influence the spectrum of the energetic particles in the shock.

This leads to the basic thesis underlying all the arguments made here: We propose that *a principle of the smallest dominant scale,* either in real space or in velocity space (Biermann 1993a), allows us to determine the relevant transport coefficients, which describe the overall transport of particles in the shocked region, both parallel and transverse to the shock direction, and to derive drift energy gains.

## D. Explosions: Interstellar medium versus Winds

One might ask what the essential difference actually is between explosions into a stellar wind and an explosion into the interstellar medium: There are three important differences, i) one is that a stellar wind has a density gradient, asymptotically of density $\rho(r) \sim r^{-2}$ with radius $r$, leading to persistent high shock velocities, ii) the second is the possibility that stellar winds have much stronger magnetic fields than the interstellar medium, and iii) that the winds of massive stars are enriched in heavy elements in the last stages of their evolution, thus strongly biasing any energetic particle population.

As a consequence, the first phase of an explosion, when the ejected mass $M_{ej}$ is in free expansion until an approximately equal amount of mass is snowplowed together, has a quite different characteristic radius $R_e$: For an interstellar medium of ion density $n_i$ this radius is

$$R_e(ISM) = 1.97 \, \text{pc} \, (\frac{M_{ej}}{M_\odot})^{1/3} \, n_i^{-1/3}. \qquad (1)$$

Usually, this radius is exceeded even for low densities of the interstellar medium, so that the phase of expansion with constant energy is relevant for any discussion, the Sedov phase. This means that the shock speed $U_1$ is steadily decreasing with radius, as $U_1 \sim r^{-3/2}$.

In contrast, the corresponding radius for the expansion into a wind is given by

$$R_\varepsilon(wind) \; = \; 306\mathrm{pc} \; \frac{M_{ej}}{M_\odot} \; \frac{V_{W,-2}}{\dot{M}_{-5}}. \tag{2}$$

Here $V_{W,-2}$ is the wind velocity in units of $10^{-2}$ c, and $\dot{M}_{-5}$ is the wind mass loss in units of $10^{-5} \, M_\odot$ per year. This radius is, obviously, reduced for the explosive expansion into a red giant wind, with a wind speed of only 30 km/sec, by a factor of 100 to only a few parsec, everything else being equal. However, even then this radius will usually not be reached or at least not be surpassed by a considerable factor.

As a result the shock velocity of an explosion into the interstellar medium is typically slowing down steadily, while in striking contrast the explosion into a wind can be expected to be in an approximately free expansion phase and so at nearly constant high velocity.

Another question, one might ask, is whether an explosion into a wind does not naturally lead to an explosion into the interstellar medium, when the edge of the wind bubble is reached. From evolutionary calculations of massive stars it is clear that there is considerable mass loss before the final explosion, and all this mass forms, together with snowplowed interstellar medium material, a thick shell around the wind zone. Hence for those stars, which have a large amount of mass loss through a wind, the supernova shock has some difficulty penetrating this shell around the wind bubble without considerable energy loss. One difference between wind supernovae and interstellar medium supernovae is then that in the first case the explosion dissipates a significant fraction of its energy in the wind bubble shell, while in the second case the explosion can break through and form a remnant in adiabatic expansion. Wheeler (1989) shows that the dividing line is likely to be near a zero-age main sequence mass of about 15 solar masses; we discuss this point below (section VI) in the context of the energy requirements of the galactic cosmic rays attributed to the two source populations.

### E. Outline

The review is based on earlier work, summarizes it and expands upon it (Biermann 1993a, paper CR I; Biermann & Cassinelli 1993, paper CR II; Biermann & Strom 1993, paper CR III; Stanev *et al.* 1993, paper CR IV; Rachen & Biermann 1993, paper UHE CR I; Rachen *et al.* 1993, paper UHE CR II; Nath & Biermann 1993, 1994a, b; Biermann, Gaisser & Stanev 1994) with earlier reviews in Biermann (1993b, 1994a, b, c) with, however, an entirely different emphasis.

Much of the new material described here is based on discussions the author had with other participants at the meetings in Raleigh, NC (September 1993, organized by D. Ellison & S. Reynolds), in Tucson, AZ (October 1993, organized by C.P. Sonett, M.S. Giampapa &

J.R. Jokipii), in Budapest, Hungary (March 1994, organized by Zs. Nemeth and E. Sormorjai), in Vulcano, Italy (May 1994, organized by F. Giovannelli and G. Mannocchi), in Nandaihe, China (August 1994, organized by Z.G. Deng, X.-Y. Xia and G. Börner), in St. Petersburg, Russia (September 1994, organized by D.A. Varshalovich, A. Bykov and others), and in Stockholm, Sweden (September 1994, organized by L. Bergström, P. Carlson, P.O. Hulth, and H. Snellman). Not all reports written for these meetings are mentioned.

In the following we will describe in section II the basic concept for particle acceleration in spherical shocks, in section III we will derive the properties of the energetic particle population, in section IV we will discuss the implications for stellar winds, in section V we will discuss the consequences for energetic electrons, in section VI for energetic nuclei, and summarize in section VII.

## II. The shock region

### A. A basic instability

The normal asymptotic configuration for an embedded magnetic field in a stellar wind has been derived by Parker (1958), and gives a magnetic field, which is tangential, decreases with radius $r$ as $1/r$, and with colatitude $\theta$ as $\sin \theta$.

Consider a spherical shock wave in such a wind. We will use the approximation throughout, that the shock wave is spherically symmetric, and that any asymmetry is introduced by the orientation and latitude dependence of the magnetic field only. This is a strong simplification, but is necessary to keep the issue clear which we discuss here, the physics of shock waves which are at least locally spherical, and which run through a stellar wind.

Then a spherical shock wave, centered on the star in its symmetry, compresses the magnetic field, this being tangential, by the full compression factor of 4 for a strong shock in an adiabatic gas of index 5/3.

Particles can be injected into an acceleration process, and give a large proportion of the overall pressure and energy density (see, *e.g.*, Ellison *et al.* 1990, Jones & Ellison 1991). Then we have the case of a cosmic ray modified shock wave, such as treated by Zank *et al.* (1990). In the unperturbed state we have here a configuration, where the magnetic field is perpendicular to the shock direction, *i.e.* parallel to the shock surface. Zank *et al.* demonstrated that for this case the configuration is neutrally stable, with an increasing instability as soon as the shock becomes more oblique relative to the underlying magnetic field, leading to maximum instability for the case of a parallel shock configuration. In the oblique shock configuration as many as three instabilities may operate (see their Table 1). In the case, which we consider here, this means that the slightest perturbation of the shock surface relative to the underlying magnetic field is leading to an instability. Such an instability then increases the deformation of the surface, which in turn strengthens the instability. Ultimately, the shock surface is maximally deformed, and changes its shape continuously, since the configuration is stuck in an unstable mode. Thus, strong turbulence is expected in the shock region, including downstream (Ratkiewicz *et al.* 1994).

This means then, for instance, that at the tip of an outward bulge of the shock surface the bulge can move outward with respect to the time averaged shock frame, until a column density is encountered equal to the column density behind the shock; then the bulge, driven by the instability, has to slow down in its local motion. At the sides of the bulge, energetic particles readily can leak into the upstream region, going through the shock just once in a locally highly oblique configuration (*i.e.* magnetic field versus shock normal), before the shock overruns them again. This is equivalent to saying, that the local acceleration efficiency is strongly reduced, because here we envisage an energetic particle to spend a fair amount of time on either side of the shock, gaining an appreciable amount of energy from drifts, before going back through the shock.

Such a picture then leads to a concept, where the shock surface is jumping around an average location, a sphere in our case. The upstream length scale of this jumping is a length corresponding to the same column density as the average downstream region, which corresponds to all the matter snowplowed in the expansion of the spherical shock. It is important to note that the length scales associated with the instability are hydrodynamic and therefore we believe to be justified to use hydrodynamic scales below; we also adopt below the gross simplification of the maximum density jump of 4 valid for a normal gas with adiabatic coefficient of 5/3 here ignoring any cosmic ray modification (see, *e.g.*, Duffy *et al.* 1994). In this concept the acceleration is a combination of a) the drifts the particles experience in the upstream or downstream regions, since the averaged magnetic fields are, of course, still perpendicular to the shock direction, and b) the energy gain from the Lorentz transformation each time a particle goes through the shock. At the same time, particles also lose energy from adiabatic expansion, since in the expansion of a spherical shock the local length scales always increase.

We have to caution, that we use here results from cosmic ray transport theory in an environment for which this theory was not made, *i.e.* where the magnetized ionized plasma is dominantly turbulent.

## B. Radiopolarization

How can we test such a concept with observations? The *radio ob-*

*servations* of young supernova remnants and the nova GK Per can be a guide here. Radio polarization observations of supernova remnants clearly indicate what the typical local structure of these shocked plasmas is. Obviously, the normal expansion of a supernova remnant is not into a stellar wind, but into the interstellar medium However, the typical magnetic field is highly oblique on average in a random field, and so the essential issue remains the same as in the case of a shock into a wind. The observational evidence (Milne 1971, Downs & Thompson 1972, Reynolds & Gilmore 1986, Milne 1987, Dickel *et al.* 1988) has been summarized by Dickel *et al.* (1991) in the statement that all shell type supernova remnants less than 1000 years old show dominant radial structure in their magnetic fields near their boundaries. There are several possible ways to explain this; we concentrate here on the idea, that this polarization pattern is due to rapid convective motion which induces locally strong shear.

These examples are for supernova explosions into the interstellar medium; there is also an observation demonstrating the same effect for an explosion into a wind: Seaquist *et al.* (1989) find for the spatially resolved shell of the nova GK Per a radially oriented magnetic field in the shell, while the overall dependence of the magnetic field on radial distance $r$ is deduced to be $1/r$ just as expected for the tightly wound up magnetic field in a wind. The interpretation given is that the shell is the higher density material behind a shock wave caused by the nova explosion in 1902 and now travelling through a wind. Seaquist *et al.* (1989) note the similarity to young supernova remnants.

The important conclusion for us here is that there appear to be strong radial differential motions in perpendicular shocks which provide the possibility that particles get *convected* parallel to the shock direction. The instability described earlier (Zank *et al.* 1990) may be the physical reason for this rapid convective motion, we would like to suggest. We assume this to be a diffusive process, and note that others have also pointed out that this may be a key to shock acceleration (*e.g.* Falle 1990).

It is important to emphasize that older supernova remnants do not show a clear pattern as described above. Already the data of the radio knot motions in Cas A (Tuffs 1986) clearly show a rather chaotic behaviour; it appears that the radio knots move erratically with a speed of the order of the shock speed itself. This is actually important, it turns out: we will use this erratic motion below to limit the drift effect along the shock sphere.

## C. Tycho versus Cygnus Loop

One important consequence of the concept introduced here is that for spherical shocks, that do not accelerate a new particle population to high energy, the overall shell thickness should be just that due to the snowplow; for a spherical shock in a homogeneous medium this is $r/12$ in the case of a strong shock. Obviously, if cooling becomes important, then the thickness is even less. For a shock that does accelerate a new energetic particle population in a shock in the interstellar medium, the instability described above leads to an estimate of the shell thickness of $5r/16$, again for a strong shock, and referred to the outer edge (see below).

The Cygnus Loop as well as Tycho appear to show indeed such properties; the Cygnus loop does not require a new particle population for its radio emission, the squeezed interstellar medium is well sufficient, and the shell is thin (Raymond 1993, pers.comm.; Green 1990); at the same time, the supernova remnant Tycho (Dickel *et al.* 1991) does require a newly accelerated particle population to explain the radio observations and its shell thickness over most of its circumference is close to prediction, *i.e.* fairly thick. Also here we note that such an agreement does not *prove* our concept to be correct, but it does show consistency. We will discuss this point at some more length below.

## III. The spectrum

### A. Derivation

The important conclusion for us here is that *observations and theoretical arguments suggest* the existence of strong radial differential motions in those perpendicular shocks which are mediated by cosmic rays; this in turn suggests that particles get *convected* parallel to the shock direction. We emphasize that convective motion at a given scale entails that particle diffusion is independent of energy. We assume this convective turbulence with associated particle transport to be a diffusive process, for which we have to derive a natural velocity and a natural length scale, which can be combined to yield a diffusion coefficient. A classical prescription is the method of Prandtl (1925): In Prandtl's argument an analogy to kinetic gas theory is used to derive a diffusion coefficient from a natural scale and a natural velocity of the system. Despite many weaknesses of this generalization Prandtl's theory has held up remarkably well in many areas of physics far beyond the original intent. In a leap of faith we will use a similar prescription here.

In order to generalize, we introduce the notion of the *smallest dominant scale*. This can be a scale in length or in velocity, may refer to an anisotropic transport, and thus be different in orthogonal directions. This *principle* does not say that nature lets convective transport compete on the basis of different scales, because then the longest scale would be the fastest transport, and would win; it rather implies that nature chooses for the effective convective transport the smallest dominant scales. The idea, that the smallest dominant scale may, *e.g.*, be related to the distance to a wall in a fluid current, has already been used for a long time (see section III §5 in Prandtl 1949). We use the assumption that the smallest dominant scale is the relevant scale several times in the course of this work in order to derive diffusion coefficients and other scalings. In some cases, when a smallest dominant scale is energy dependent, there may be a switch from one physical scale to another, say, to an energy-independent scale. Such a switch defines critical particle energies.

Consider the structure of a layer shocked by a Supernova explosion into a stellar wind in the case, that the adiabatic index of the gas is 5/3 and the shock is strong. Then there is an inherent length scale in the system, namely the thickness of the shocked layer, in the spherical case for a shock velocity much larger than the wind speed and in the strong shock limit $r/4$. This is the thickness of the matter snowplowed all the way from the star to the current location of the shockfront, in the simplified picture of constant density throughout the shell. There is also a natural velocity scale, namely the velocity difference of the flow with respect to the two sides of the shock. Both are the smallest dominant scale, in velocity and in length.

Our basic conjecture, **argument 1**, *based on observational evidence as well as theoretical arguments*, is then that the convective random walk of energetic particles perpendicular to the unperturbed magnetic field can be described by a diffusive process with a downstream diffusion coefficient $\kappa_{rr,2}$ which is given by the thickness of the shocked layer and the velocity difference across the shock, and is independent of energy:

$$\kappa_{rr,2} \;=\; \frac{1}{3}\,\frac{U_2}{U_1}\,r\,(U_1 - U_2) \tag{3}$$

This is in apparent contradiction to the normal argument that (for relativistic particles) $\kappa_{rr} = c\,\lambda(E)/3 > c\,r_g/3$, where $\lambda(E)$ is the mean free path for resonant scattering of a particle of energy $E$ and Larmor radius $r_g$. We do not invoke resonant scattering, but fast convective turbulence as the dominant process. We do not ignore resonant scattering, but use Jokipii's argument (1987) on permissible scattering coefficients in oblique geometries (see just below) to demonstrate that we are within bounds here.

The upstream diffusion coefficient can be derived in a similar way, but with a larger scale. We make here the second critical step, **argument 2**, namely that the upstream length scale is just $U_1/U_2$ times larger, and so is $r$. This is the relevant scale for the same column density on both sides of the average shock location, and can be argued on the basis of what limits the instability (see above). This, also, is the same ratio as the mass density and the ratio of the gyroradii of the same particle energy. Since the magnetic field is lower by a factor of $U_1/U_2$ upstream, that means that the upstream gyroradius of the maximum energy particle that could be contained in the shocked layer, is also $r$. Hence the natural scale is just $r$. And so the upstream diffusion coefficient is

$$\kappa_{rr,1} = \frac{1}{3} r (U_1 - U_2)$$ (4)

It immediately follows that

$$\frac{\kappa_{rr,1}}{r U_1} = \frac{\kappa_{rr,2}}{r U_2} = \frac{1}{3} (1 - \frac{U_2}{U_1}).$$ (5)

For these diffusion coefficients, it also follows that the residence times (Drury 1983) on both sides of the shock are equal and are

$$\frac{4 \kappa_{rr,1}}{U_1 c} = \frac{4 \kappa_{rr,2}}{U_2 c} = \frac{4}{3} \frac{r}{c} (1 - \frac{U_2}{U_1}).$$ (6)

Here it has to be stated that the residence time is normally derived from a diffusion argument which is obviously not valid here, because it requires that the scattering length is much smaller than the dominant scale, while we basically identify those two scales. We have not proven, but surmise that the concept of the residence time could be rederived from a time-averaging of the probability that a given particle is still on the same side of the jumping and wobbling shock surface after some time $t$. It is an assumption, based on dimensional arguments, that the result would be the same.

Adiabatic losses then cannot limit the energy reached by any particle since they run directly with the acceleration time, both being independent of energy, and so the limiting size of the shocked layer limits the energy that can be reached to that where the gyroradius just equals the thickness of the shocked layer, provided the particles can reach this energy. We assume here that the average of the magnetic field $\langle B \rangle$ is not changed very much by all this convective motion. This then leads to a maximum energy of

$$E_{max} = \frac{U_2}{U_1} Z e r B_2 = Z e r B_1$$ (7)

where $Ze$ is the particle charge and $B_{1,2}$ is the magnetic field strength on the two sides of the shock. This means, that the energy reached corresponds to the maximum gyroradius the system will allow on both sides of the shock. It also says that we push the diffusive picture right up its limit where on the downstream side the diffusive scale becomes equal to the mean free path and the gyroradius of the most energetic particles.

Jokipii (1987) has derived a general condition for possible values of the diffusion coefficient: Its value has to be larger than the gyroradius multiplied by the shock speed. This condition is fulfilled here, for the maximum energy particles only by a factor of $1 - U_2/U_1 < 1$, since here the shock speed and the radial scale of the system give both the largest gyroradius as well as the diffusion coefficient. This also counters part of the criticism of Völk in Ellison *et al.* (1994) - the other part of his criticism is dealt with in section 11 of paper CR I.

Observations can now give information of possible values for the diffusion coefficient as well: Smith *et al.* (1994) estimate the cosmic ray diffusion coefficient at shocks in the LMC and find severe upper limits. However, first of all, they use the *assumption* that the underlying unperturbed magnetic field is parallel to the shock normal; second, they measure an instantaneous situation, whereas I have argued above that strong and fast convective motion dominates the shock region. The diffusion coefficient derived above is an effective temporal and geometric average over this fast convection. Hence there is no contradiction.

There is an important consequence of this picture for the diffusion laterally: From the residence timescale and the velocity difference across the shock we find a distance which can be traversed in this time of

$$\frac{4}{3}\,\frac{r}{c}\,(1 - \frac{U_2}{U_1})\,(U_1 - U_2). \tag{8}$$

Since the convective turbulence in the radial direction also induces motion in the other two directions, with maximum velocity differences of again $U_1 - U_2$, this distance is also the the typical lateral length scale. We noted above that the observed motions of radio knots in the supernova remnant Cas A support such an argument (Tuffs 1986). From this scale and again the residence time we can construct an upper limit to the diffusion coefficient in lateral directions of

$$\kappa_{\theta\theta,max} = \frac{4}{9}\,(1 - \frac{U_2}{U_1})^3\,(\frac{U_1}{c})^2\,r\,c, \tag{9}$$

which is for strong shocks equal to

$$\kappa_{\theta\theta,max} = \frac{1}{3}\,(\frac{3}{4}\,\frac{U_1}{c})^2\,r\,c. \tag{10}$$

Again in the spirit of the idea, that the smallest dominant scale determines the effective transport, this then will begin to dominate as soon as the $\theta$-diffusion coefficient reaches this maximum at a critical energy. As long as the $\theta$-diffusion coefficient is smaller, it will dominate particle transport in $\theta$ and the upper limit derived here is irrelevant. When the $\theta$-diffusion coefficient reaches and passes this maximum, then the particle in its drift will no longer see an increased curvature due to the convective turbulence because of averaging and the part (here we have to account for both losses and gains by drifts) of drift acceleration due to increased curvature is eliminated. This then reduces the energy gain, and the spectrum becomes steeper from that energy on. The critical particle energy thus implied will be identified below with the particle energy at the knee of the observed cosmic ray spectrum.

## A.1. Drifts

Consider particles which are either upstream of the shock, or downstream; as long as the gyrocenter is upstream we will consider the particle to be there, and similarly downstream.

In general, the energy gain of the particles will be governed primarily by their adiabatic motion in the electric and magnetic fields. The expression for the energy gain is then (Northrop, 1963, eq. 1.79), for an isotropic angular distribution

$$\frac{dE}{dt} = Ze\,\mathbf{V_d}\frac{\mathbf{U} \times \mathbf{B}}{c}\bigg|_i + \frac{pw}{c}\frac{\partial lnB}{\partial t} \tag{11}$$

where the first term arises from the drifts and the second from the induced electric field. This equation is valid in any coordinate frame. We explicitly work in the shock frame, and the separate the two terms above and consider the drift term first. The second term is accounted for further below.

The $\theta$-drift velocity in a normal stellar wind is given above. The $\theta$-drift can be understood as arising from the asymmetric component of the diffusion tensor, the $\theta r$-component. The natural scales there are the gyroradius and the speed of light, and so we note that for (Forman *et al.* 1974)

$$\kappa_{\theta r} = \frac{1}{3}\,r_g\,c, \tag{12}$$

the exact limiting form derived from ensemble averaging, we obtain the drift velocity by taking the proper covariant divergence (Jokipii *et al.* 1977); this is not simply (spherical coordinates) the $r$-derivative of $\kappa_{\theta r}$. The general drift velocity is given by (see, *e.g.*, Jokipii 1987)

$$V_{d,\theta} = c\,\frac{E}{3Ze}\,curl_\theta\frac{\mathbf{B}}{B^2}. \tag{13}$$

The $\theta$-drift velocity is thus

$$V_{d,\theta} = \frac{2}{3}\,c\,r_g/r, \tag{14}$$

where $r_g$ is now taken to be positive. This drift velocity is just that due to the gradient as well as the curvature, and in fact both effects contribute here equally.

It must be remembered that there is a lot of convective turbulence which increases the curvature: The characteristic scale of the turbulence is $r/4$ for strong shocks, and thus the curvature is $4/r$ maximum. Here we have to consider the balance of drift energy gains and losses: We call the motion of a convective element that moves in the upstream direction upflow, and for a downstream direction of motion, we call it downflow. Then the direction of an upflow convective element is not in general just radial, and so we have to take the average of an assumed omnidirectional distribution of upflow directions, weighting it with the probability of that particular direction, and obtain $2/3$ of the maximum curvature (same averaging as done in eq. 2.47 of Drury [1983] for the averaging of the momentum gain). It is plausible that for upflow directions we have to take the maximum curvature, since the velocity difference to the surroundings is maximum. Downflow, however, the velocity difference to the surroundings is only $1/4$ of that upflow, and so we take for the downflow convection a curvature of $1/4$ of the maximum, multiplied also by $2/3$. The difference of gains and losses is then the net gain, thus giving a factor of $(2/3)(1 - 1/4) = 1/2$ of the maximum curvature (**argument 3**). This is equivalent to taking half the maximum as average for the net energy gain due to drifts. We obtain then for the curvature a factor of $2/r$ which is twice the curvature without any turbulence; this increases the curvature term by a factor of two thus changing its contribution from $1/3$ to $2/3$ in the numerical factor in the expression above. Hence the total drift velocity, combining now again the curvature $(2/3)$ and gradient $(1/3)$ terms, is thus

$$V_{d,\theta} \;=\; \frac{1}{3}\,(1 + \frac{U_1}{2U_2})\,c\,r_g/r, \qquad (15)$$

now written for arbitrary shock strength. It is easily verified that the factor in front is unity for strong shocks where $U_1/U_2 = 4$.

The energy gain associated with such a drift is given by the product of the drift velocity, the residence time, and the electric field. Upstream this energy gain is given by

$$\Delta E_1 \;=\; \frac{4}{3}\,E\,\frac{U_1}{c}\,f_d\,(1 - \frac{U_2}{U_1}), \qquad (16)$$

where

$$f_d \;=\; \frac{1}{3}\,(1 + \frac{U_1}{2U_2}). \qquad (17)$$

Thus, $f_d = 1$ for strong shocks. The corresponding expression downstream is

$$\Delta E_2 \;==\; \frac{4}{3}\,E\,\frac{U_2}{c}\,f_d\,(1 - \frac{U_2}{U_1}), \qquad (18)$$

giving a total energy gain of

$$\Delta E/E = \frac{4}{3}\,\frac{U_1}{c}\,f_d\,(1+\frac{U_2}{U_1})\,(1-\frac{U_2}{U_1}). \tag{19}$$

The drift energy gain averages over the magnetic field strength during the gyromotion. We emphasize that this energy gain is independent of this average magnetic field, so that even variations of the magnetic field strength due to convective motions do not change this energy gain.

It can easily be shown, that this treatment can be extended to sub-relativistic energies, and provides powerlaws in momentum as a result, just as in standard case (see below, and Drury 1983).

It is of interest to note here, that the net distance travelled (i.e. drifted) by the particle, e.g. upstream, is given by

$$l_{\perp 1} = \frac{4}{3}\,\frac{E}{ZeB_1}\,f_d\,(1-\frac{U_2}{U_1}) \tag{20}$$

which is the gyroradius itself for $U_2/U_1 = 1/4$, corresponding to a strong shock. This then says that we are at a gyroradius limit for the drift distance. Just as in isotropic turbulence the gyroradius is a lower limit to the mean free path for particle scattering parallel to the magnetic field in a turbulent plasma, suggesting that it may be useful to think of the plasma also as maximally turbulent perpendicular both to the flow and to the magnetic field. We emphasize that during this drift the particle makes many gyromotions. It is also important to note that the magnetic field structure in the shocked region - as discussed above on the basis of observations - will contain local regions of opposite magnetic field and so the drift itself will be erratic and be the sum of many single element drift movements. What we have derived is the average net energy gain due to drifts, with the drift distance corresponding to the average magnetic field strength.

## A.2. The energy gain of particles

Shock acceleration in its standard form just uses the Lorentz transformation for an energetic particle at a velocity $v$ much larger than the shock velocity to compute the energy gain as the particle goes between scattering in a weakly turbulent magnetic field from downstream to upstream and back. In practice we will consider the case when $v$ is close to $c$.

We assume the shock to be subrelativistic and so the phase space distribution of the particles to be nearly isotropic. Then downstream (see Drury 1983 for the exact derivation) the particles have a finite chance to escape. A detailed discussion yields a powerlaw for the distribution $p^2 f(p)$, where $f(p)$ is the particle distribution function of momentum $p$ in phase space, with the powerlaw index -4 for strong shocks for a gas of adiabatic index 5/3 (i.e. $p^2 f(p) \sim p^{-2}$).

However, this assumes that there are no other energy losses or energy gains during a cycle of a particle going back and forth between upstream and downstream. In a configuration, where there is a component of the magnetic field perpendicular to the shock normal, there can be energy gains by drifts parallel to the electric field seen by a particle moving with the shock system, and also losses due to adiabatic expansion in the expansion of a curved shock. Since drift acceleration is a rate, the resulting particle energy gain is proportional to the time spent on either side of the shock. Furthermore, there can be spectral changes due to the fact, that particles were injected at a different rate in the past, when some particle under consideration now was injected; again, this is likely to happen in a spherical shock.

Let us consider then one full cycle of a particle remaining near the shock and cycling back and forth from upstream to downstream and back. The energy gain just due to the Lorentz transformations in one cycle can then be written as

$$\frac{\Delta E}{E}_{LT} = \frac{4}{3}\frac{U_1}{c}(1 - \frac{U_2}{U_1}). \tag{21}$$

Adding the energy gain due to drifts we obtain

$$\frac{\Delta E}{E} = \frac{4}{3}\frac{U_1}{c}(1 - \frac{U_2}{U_1})x, \tag{22}$$

where

$$x = 1 + \frac{1}{3}\left(1 + \frac{U_1}{2U_2}\right)\left(1 + \frac{U_2}{U_1}\right), \tag{23}$$

which is $9/4$ for a strong shock, when $U_1/U_2 = 4$.

It is easy to show that the additional energy gain flattens the particle spectrum by

$$\frac{3\,U_2}{U_1 - U_2}\left(1 - \frac{1}{x}\right). \tag{24}$$

### A.3. Expansion and injection history

Consider how long it takes a particle to reach a certain energy:

$$\frac{dt}{dE} = \{8\,\frac{\kappa_{rr,1}}{U_1 c}\}/\{\frac{4}{3}\frac{U_1}{c}(1 - \frac{U_2}{U_1})xE\}. \tag{25}$$

Here we have used that $\kappa_{rr,1}/U_1 = \kappa_{rr,2}/U_2$. Since we have

$$r = U_1 t \tag{26}$$

this leads to

$$\frac{dt}{t} \;=\; \frac{dE}{E}\,\frac{3U_1}{U_1 - U_2}\,\frac{2}{x}\,\frac{\kappa_{rr,1}}{rU_1} \tag{27}$$

and so to a dependence of

$$t(E) \;=\; t_o\,(\frac{E}{E_o})^\beta \tag{28}$$

with

$$\beta \;=\; \frac{3U_1}{U_1 - U_2}\,\frac{2}{x}\,\frac{\kappa_{rr,1}}{rU_1} \tag{29}$$

which is a constant independent of $r$ and $t$.

Particles that were injected some time ago were injected at a different rate, say, proportional to $r^b$. This then leads to a correction factor for the abundance of

$$(\frac{E}{E_o})^{-b\beta}. \tag{30}$$

However, in a $d$-dimensional space, particles have $r^d$ more space available to them than when they were injected, and so we have another correction factor which is

$$(\frac{E}{E_o})^{-d\beta}. \tag{31}$$

The combined effect is a spectral change by

$$-\frac{3U_1}{U_1 - U_2}\,\frac{2}{x}(d+b)\frac{\kappa_{rr,1}}{rU_1}. \tag{32}$$

Thus we have a density correction factor, which depends on the particle energy, and so changes the spectrum.

This expression can be compared with a limiting expansion derived by Drury (1983; eq. 3.58), who also allowed for a velocity field; Drury (1983) generalized earlier work on spherical shocks by Krymskii & Petukhov (1980) and Prishchep & Ptuskin (1981). Drury's expression agrees with the more generally derived expression given here for $x = 1$. The comparison with Drury's work clarifies that for $\kappa \sim r$ the inherent time dependence drops out except, obviously, for the highest energy particles, discussed further below; the same comparison shows that the statistics of the process are properly taken into account in our simplified treatment. We note that $\frac{\kappa_{rr,1}}{rU_1}$ is 1/4 in the wind case, and 1/12 in the ISM case, both for strong shocks, and thus still small compared to unity. A fortiori, the comparison shows that in this limit of small $\frac{\kappa_{rr,1}}{rU_1}$, the derivation by Drury, using the properly derived classical cosmic ray transport theory, and our heuristic derivation, both agree.

This agreement does not provide proof of our microscopic picture, but it is an important check.

If the expansion is linear, as it is the case here, then the $r^d$-term also describes the adiabatic losses in their effect on the spectrum, due to the general expansion of the shock layer and thus accounts for the second term in eq. (11) above. Hence the total spectral difference, as compared with the planeparallel case, is given by

$$\frac{3U_1}{U_1 - U_2} \{ \frac{U_2}{U_1}(\frac{1}{x} - 1) \ + \ \frac{2}{x}(b+d)\frac{\kappa_{rr,1}}{rU_1} \}. \qquad (33)$$

Here we use the following sign convention: For this expression positive, the spectral index of the particle distribution is steeper than without this correction; this then takes the minus sign in eqs. (30 - 32) properly into account.

This expression (33) constitutes the basic result of this section. For a wind we have $b = -2$ and $d = +3$, and so $b + d = 1$. For a Sedov-type expansion into a homogeneous medium the additional differential adiabatic losses steepen the spectrum further (paper CR III). The total spectral change is then for $U_1/U_2 = 4$ given by 1/3, so that the spectrum obtained is

$$\text{Spectrum (source)} \ = \ E^{-7/3}. \qquad (34)$$

This is what we wanted to derive. Generalizing to transrelativistic energies, we note that the spectrum is a powerlaw in relativistic momentum $p$, as can easily be seen from the detailed derivation.

Generalizing now for arbitrary wind speed and arbitrary shock strength we obtain a reduced thickness of the shocked layer. This reduction of the thickness of the layer for a finite wind velocity is due to the fact that the material which is snowplowed together is not all gas between zero radius and the current radius $r$, but between zero radius and $r(1 - V_W/(V_W + U_1))$, since the gas keeps moving while the shock moves out towards $r$. For the sequence of $V_W/U_1 = 0., 1.0$, and $\gg 1$ we thus obtain particle spectral index differences, in addition to the index of 7/3, of 0.0, 0.136, 0.303, corresponding to synchrotron emission spectral index of an electron population with the same spectrum, of $-0.667, -0.735, -0.818$ (using the convention that flux density $S_\nu \sim \nu^\alpha$). The work of Owocki et al. (1988) suggests that typical shocks in winds have a velocity in the wind frame similar to the wind velocity in the observers frame itself, which implies that, in the simplified picture here, only spectral indices for the synchrotron emission between $-0.667$ and $-0.735$ are relevant, with an extreme range of spectral indices up to $-0.818$ for strong shocks. Obviously, for weaker shocks with $U_1/U_2 < 4$ the spectrum can be steeper, e.g. for $U_1/U_2 = 3.5$ we obtain an optically thin spectral index for the synchrotron emission of $-0.734$ for $V_W \ll U_1$ and $-0.815$ for $V_W/U_1 = 1$. This then is the prediction for the spectral index of the nonthermal radio emission from massive stars like OB stars and Wolf Rayet stars, as well as for other shocks in winds like around novae.

We can estimate the uncertainty only with difficulty, since the argument is a limit, the limit of a strong shock for instance, and also in all other steps of the argument we have used the conceptual limit of the scales involved. However, we can estimate one uncertainty, which arises from the finite wind speed of Wolf Rayet stars, or similar stars with strong winds which explode as supernovae. These wind speeds can go up to several thousand km/sec, while the supernova shock is variously estimated to $10^4$ km/sec to twice that much. As a limiting argument we use that the ratio of the wind speed to the supernova shock speed is $< 0.2$; this gives a steepening of the derived spectral index of the particle distribution by $0.04$. This uncertainty also may correspond to curvature of the spectrum, since there is a time-evolution as the shock progresses out through the stellar wind: As more energy of the shock is dissipated and more mass of the stellar wind snow-plowed, the shock slows down; then those particles already accelerated keep their flatter spectrum (see eq. 2.44 of Drury 1983), while those particles freshly injected and accelerated will have a steeper spectrum. Thus, in the range $V_W/U_1 = 0., ..., 0.2$ we obtain a spectral index in the range $7/3, ..., 7/3 + 0.04$. We note that the evolved stars which are most common have a slow wind, and so the distribution of stars through this adopted range of wind velocities is likely to be biased towards the small numbers. Therefore we ascribe to the spectral index derived here an uncertainty of $-0.02 \pm 0.02$, which describes both the uncertainty in an assumed powerlaw, and the possible curvature. After correcting for leakage from the Galaxy the spectrum is

$$\text{Spectrum (earth)} = E^{-8/3 - 0.02 \pm 0.02} \qquad (35)$$

very close to what is observed near earth at particle energies below the knee. We plan to discuss the error estimates in a separate contribution.

It is of interest to note, that an injection spectrum of $E^{-7/3}$ can also lead via pp-collisions in a synchrotron dominated regime to a spectrum of pair-secondaries of $E^{-10/3}$, which translates in the synchrotron spectrum to a $-7/6$ powerlaw flux density spectrum and a $-13/6$ powerlaw photon number spectrum, very close to that observed by GRO for the Crab pulsar (Schönfelder 1992, seminar in Bonn). If the curvature of a shock is not exactly spherical, obviously, some differences to these particular numerical values will occur. Such a speculative interpretation would place the origin of the pulses in periodically excited shocks travelling down a perpendicular magnetic field configuration in a pulsar wind as considered here.

Such an injection spectrum of $-7/3$ of relativistic particles in strong and fast shocks propagating through a stellar wind leads to an unambiguous radio synchrotron emission spectrum of $\nu^{-2/3}$ in flux density $S_\nu$ (compare, *e.g.*, the nonthermal radioemission of OB stars, WR stars, novae, especially GK Per: Reynolds & Chevalier 1983, 1984, radio supernovae, and supernova remnants). The observed variety of spectral indices of the radio emission of supernova remnants is discussed in some detail in paper CR III.

## B. The maximum energy of particles

The maximum energy particles can reach is given above, and depends linearly on the magnetic field. Thus, we require estimates for the magnetic field in the stellar winds of Wolf Rayet stars and other massive stars, that explode as Supernovae, like red and blue supergiants. Comparing at first the corresponding estimates that Völk & Biermann (1988) used, we note that the energies implied here are larger by approximately $c/U_1$ for the same given magnetic field strength, since their expression for the maximum energy that particles could reach contains an additional factor of approximately $U_1/c \ll 1$ as compared with our eq. (7). This modified limit of Völk & Biermann to the particle energy, which can be reached, corresponds well to the limit valid for supernova explosions into the interstellar medium (see paper CR III).

Cassinelli (1982), Maheswaran & Cassinelli (1988, 1992) have argued that Wolf Rayet stars have very much larger magnetic fields than Völk & Biermann used, in order to drive their winds. The magnetic fields given by Cassinelli and coworkers are of order a few thousand Gauss on the surface of the star. We introduce the conjecture here, discussed in more detail in paper CR II and below, that the Alfvén radius of the stellar wind is close to the stellar surface itself. Then it follows that the product $Br$ has approximately the same value on the surface as in the wind, and is of order $3\,10^{14}\,B_{0.5}\,\mathrm{cm\,Gauss}$, where $B_{0.5}$ is the strength of the magnetic field at $10^{14}$ cm radial distance in units of 3 Gauss. From this number we infer a maximum energy of particles of

$$E_{max}(\mathrm{protons}) \;=\; 9\,10^7\,B_{0.5}\,\mathrm{GeV} \qquad (36)$$

and

$$E_{max}(\mathrm{iron}) \;=\; 3\,10^9\,B_{0.5}\,\mathrm{GeV}. \qquad (37)$$

This suggests that the highest energy particles from the acceleration process discussed here are mostly iron or other heavy nuclei. The chemical composition is expected to change abruptly to mostly protons again when the extragalactic component takes over (paper UHE CR I) somewhere near $3\,10^9$ GeV. The dependence of the maximum particle energy on the magnetic properties of the stellar winds implies a smearing due to the distribution of magnetic field strengths from all the different stars which contribute, and hence leads to a steepening of the spectrum, possibly well before the cutoff. If this mechanism provides the largest particles energies, then obviously other contributions are not excluded, by pulsars, neutron star binaries, or even from a hypothetical termination of the galactic wind.

## C. The knee in the Cosmic Ray spectrum

We wish to discuss here the bend in the spectrum of Cosmic Rays at the knee, near $5 \ 10^6$ GeV.

Let us consider the structure of the wind through which the supernova shock is running. The maximum energy a particle can reach is proportional to $\sin^2\theta$, since the space available for the gyromotion from a particular latitude is limited in the direction of the pole by the axis of symmetry. Hence, the maximum energy attainable is lowest near the poles. Then, consider the pole region itself, where the radial dependence of the magnetic field is $1/r^2$, and the magnetic field is mostly radial. We can make two arguments here: Either we put the upstream diffusive scale $4 \, \kappa_{rr,1}/(c \, U_1)$ equal to $r/c$ in the strong shock limit, or we can put acceleration time and flow time equal to each other. Both arguments lead to the same result. We use here the Bohm limit in the diffusion coefficient $\kappa_{rr,1} = \frac{1}{3}cE/Z\epsilon B(r)$, since we have a shock configuration near the pole, where the direction of propagation of the shock is *parallel* to the magnetic field - often referred to as a parallel shock configuration. This then leads to a maximum energy for the particles of

$$E \ = \ \frac{3}{4} Z\epsilon B(r) r \frac{U_1}{c},  \tag{38}$$

which is proportional to $1/r$ near the pole, where the magnetic field is parallel to the direction of shock propagation; the corresponding gyroradius is then given by $\frac{3}{4}\frac{U_1}{c}r$. Putting this equal to the gyroradius of particles that are accelerated further out at some colatitude $\theta$, where the magnetic field is nearly perpendicular to the direction of shock propagation, gives the limit where the latitude-dependent acceleration breaks down. This then gives the critical angle as

$$\sin \theta_{crit} \ = \ \frac{3}{4} \frac{U_1}{c}.  \tag{39}$$

The angular range of $\theta < \theta_{crit}$ we refer to as the *polar cap* below. We note that this angle is usually much larger than the angle where $B_r = B_\phi$ at a given radial distance $r$. If, as observed in the solar wind, the equatorial sheet of the magnetic field structure oscillates around the geometric symmetry plane, then this critical angle may either be effectively larger or the argument may fail altogether. The energy at
the location of the critical angle as defined above is then given by

$$E_{knee} = ZeB(r)r(\frac{3}{4}\frac{U_1}{c})^2. \qquad (40)$$

We identify this energy with the knee feature in the Cosmic Ray
spectrum, since all latitudes outside the polar cap contribute the same
spectrum up to this energy; from this energy to higher particle energies
a smaller part of the hemisphere contributes and also, the energy gain
is reduced, as argued below; this reduction of the energy gain in each
cycle of acceleration happens at the same critical energy. This is valid
in the region where the magnetic field is nearly perpendicular to the
shock, and thus this knee energy is independent of radius.

All this immediately implies that the chemical composition at the
knee changes so, that the gyroradius of the particles at the spectral
break is the same, implying that the different nuclei break off in order
of their charge $Z$, considered as particles of a certain energy (and not
as energy per nucleon). In a spectrum in energy per particle, this
introduces a considerable smearing.

In the polar cap the acceleration is a continuous mix between the
regime where the diffusion coefficient is determined by the thickness of
the shell, and the regime where it is dominated by turbulence parallel to
the magnetic field; this latter regime is rather small in angular extent.
Thus, $\kappa_{rr,1}/rU_1$ might be quite a bit smaller than $1/4$. Hence the polar
cap will have a spectrum which is determined by a range of

$$0 < \frac{\kappa_{rr,1}}{rU_1} < \frac{1}{3}(1 - \frac{U_2}{U_1}), \qquad (41)$$

as well as by a rather reduced role for the extra energy gain due to
drifts. Thus the spectrum is harder in the polar cap region, because
we are close to the standard parallel shock configuration, for which the
particle spectrum is well approximated by $E^{-2}$. On the other hand,
the polar cap is small relative to $4\pi$ with about $(U_2/U_1)^2$ and only a
spectrum much flatter than $E^{-7/3}$ like, $e.g.$, indeed $E^{-2}$ will make it
possible for the polar cap to contribute appreciably near the knee en-
ergy, because then the spectral flux near the knee is increased relative
to 1 GeV by $(E_{knee}/m_p c^2)^{1/3}$ which approximately compensates for its
small area. The uncertainty in the spectrum of the polar cap compo-
nent particles is not determinable at present; should the spectrum be
significantly steeper than $E^{-2}$, then this component almost certainly
is altogether insignificant. Therefore we do not ascribe any uncertainty
to this spectral index; it has to remain an assumption, derived from the
limiting argument. During an episode with drift towards the poles, a
larger part of the sphere can contribute for larger energy particles, and
so there is an additional tendency to flatten the spectrum of the polar cap contribution. The combination of the polar cap with the rest of the stellar hemisphere might lead to a situation where up to, say, $600\,Z$ TeV the entire hemisphere excluding the polar cap dominates, while from $600\,Z$ TeV up to the knee the polar cap begins to contribute appreciably. Near the knee energy the polar caps might thus contribute equally to the rest of the $4\pi$ steradians. Because of spatial limitations most of the hemisphere has to dominate again above the knee, although with a fraction of the hemisphere that decreases with particle energy. This introduces a weak progressive steepening of the spectrum with energy, which we will discuss elsewhere. The superposition of such spectra for different chemical elements, including the polar cap contribution, has been tested successfully (see paper CR IV), and is briefly summarized below, in section VI. The results of these checks suggest that the polar cap may be a reason for the flattening of the cosmic ray spectrum as one approaches the knee feature.

We note that we are using a limiting argument to derive the spectrum below the knee, and again use a limiting argument (see below) for the spectrum above the knee. Close to the knee, such an argument breaks down on either side, and so a softening of the knee feature is to be expected. On top of such a softened knee feature the polar cap gives an additional component.

The expression for the particle energy at the knee also implies by the observed relative sharpness of the break of the spectrum that the actual values of the combination $B(r)rU_1^2$ must be very nearly the same for all supernovae that contribute appreciably in this energy range; however, it is difficult to put a numerical value to this argument (see paper CR IV for the systematics that go into any parameter estimate from fitting the airshower data). Please note that $B(r)r$ is evaluated in the Parker regime, and so is related to the surface magnetic field by $B(r)r = B_s\,r_s^2\Omega_s/V_W$, where the values with index $s$ refer to the surface of the star and $V_W$ is the wind velocity. Thus, in our picture, the expression

$$B_s\,r_s^2\,\frac{\Omega_s}{V_W}\,U_1^2 \qquad (42)$$

is approximately a universal constant for all stars that explode as supernova after a Wolf Rayet phase. It may also hold for all massive stars of lower mass that explode as supernovae.

This then implies that we may have identified a functional relationship for the mechanical energy of exploding stars connecting the magnetic field, the angular momentum, and the ejection energy. Such a relationship could be fortuitous, since all massive stars become very similar to each other near the end of their evolution. But it could also be an indication for an underlying physical cause. Related ideas have been expressed and discussed by Kardashev (1970), Bisnovatyi-Kogan (1970), LeBlanc & Wilson (1970), Ostriker & Gunn (1971), Amnuel *et al.* (1972), Bisnovatyi-Kogan *et al.* (1976), and Kundt (1976), with Bisnovatyi-Kogan (1970) the closest to the argument below. All this leads to the following interesting suggestion (see paper CR I):

The source of the mechanical energy observable in supernova explosions may then be the gravitational energy of a core accretion disk at a scale determined by angular momentum and mediated by the magnetic field. We leave a discussion how this argument may lead to a surface magnetic field strength to elsewhere.

The distribution in the knee particle energy from the different stellar properties and explosive energies clearly leads to a rounding of the effective average knee shape, which softens the powerlaws on the both sides of the knee (see Biermann 1994d).

### D. The latitude distribution of the particles

Consider the derivation of the spectrum beyond the knee. Since the maximum energy particles can attain is a strong function of colatitude, the spectrum beyond the knee requires a discussion of the latitude distribution, which we have to derive first. The latitude distribution is established by the drift of particles which builds up a gradient which in turn leads to diffusion down the gradient. Hence it is clear that drifts towards the equator lead to higher particle densities near the equator, and drifts towards the poles lead to higher particle densities there. Thus the equilibrium latitude distribution is given by the balancing of the $\theta$-diffusion and the $\theta$-drift.

The diffusion tensor component $\kappa_{\theta\theta}$ can be derived similar to our heuristic derivation of the radial diffusion term $\kappa_{rr}$, again by using the smallest dominant scales. The characteristic velocity of particles in $\theta$ is given by the erratic part of the drifting, corresponding to spatial elements of different magnetic field direction. This is on average the value of the drift velocity $\mid V_{d,\theta} \mid$, possibly modified by the locally increased values of the magnetic field strength. The characteristic distance is the distance to the symmetry axis $r \sin\theta$; this is the smallest dominant scale as soon as the thickness of the shocked layer is larger than the distance to the symmetry axis, *i.e.* $\sin\theta < U_2/U_1$. Thus we can write in this approximation, **argument 4**,

$$\kappa_{\theta\theta,1} \;=\; \frac{1}{3}\;\mid V_{d,\theta} \mid\; r\,(1-\mu^2)^{1/2}. \tag{43}$$

Here $\mu$ is the cosine of the colatitude on the sphere we consider for the shock in the wind. Interestingly, this can also be written in the form

$$\frac{1}{3}\,r_g\,c\,(1-\mu^2)^{1/2},$$

where $r_g$ is taken as positive; we also note that $c\,(1-\mu^2)^{1/2}$ is the maximum drift speed at a given latitude, valid for the local maximum particle energy. This suggests that the latitude diffusion might be usefully thought of as diffusion with a length scale of the gyroradius, and the particle speed, to within the angle dependent factor, which just cancels out the latitude dependence of the magnetic field strength in the denominator of the gyroradius.

We assume then for the colatitude dependence a powerlaw $(1-\mu^2)^{-a}$ and first match the latitude dependence of the diffusion term and the drift, and then use the numerical coefficients to determine the exponent in this law. The diffusion term and the drift term have the same colatitude dependence since the double derivative and the internal factor of $(1-\mu^2)$ lead to a $(1-\mu^2)^{-a-1}$ for the diffusive term, while the drift term is just the simple derivative giving the same expression. For $(1-\mu^2) \ll 1$ the condition then is $\frac{2}{3}\,a^2 = \pm a$. The diffusive term is always positive, while the $\theta$-drift term is negative for $Z\,B_s$ negative. This means for positive particles and a magnetic field directed inwards the $\theta$-drift is towards the pole. In that case then the exponent $a$ is either zero or $a = 3/2$. Since the drift itself clearly produces a gradient, the case with $a = 0$ is of no interest here. It follows that for positive particles and an inwardly directed magnetic field the latitude distribution is strongly biased towards the poles, emphasizing in its integral the lower energies, and thus making the overall spectrum steeper beyond the knee energy; such a configuration may influence the polar cap structure in terms of magnetic fields and energetic particles. The radial drift in this case is directed outwards, which means that particles drift ahead of the shock by a small amount only to be caught up again by the diffusive region ahead of the shock. For the magnetic field directed outwards and positive particles the radial drift is inwards, taking particles out of the system at an slightly increased rate and thus steepening the overall spectrum by a small amount.

When the magnetic field is directed outwards and the particles have a positive charge, the drift is towards the equator with then a positive gradient with $(1-\mu^2)^{3/2}$, again in the limit $(1-\mu^2) \ll 1$.

We note that this exponent $3/2$ is reduced in the case, that the erratic part of the drift is increased over the steady net drift component.

One important consequence exists for the resulting radio emission: When the density of energetic particles is largest near the equator, where the magnetic field is strongest, then we have the case of maximum radio emission. In the other extreme case, that the density of energetic particles is maximal near the poles, where the magnetic field is minimal, then the radio emission is minimal, $i.e.$ unobservable usually. All this is relevant, of course, only in the simplified picture of a homogeneous structure of the stellar wind. If there were a rapid cycling of polarities, $e.g.$ much faster than on the Sun, then the period may interfere with this picture; the consequences have not been worked out yet.

## E. The spectrum beyond the knee

This means that from $E_{knee}$ the energy gain of all particles in the entire colatitude range, that is affected by diffusion, has to be considered together.

In our model for the diffusion in $\theta$ we have used the drift velocity and the distance to the symmetry axis as natural scales in velocity and in length. When the $\theta$-drift reaches the maximum derived earlier, eq. (10), then the latitude drift changes character. This happens at a critical energy, which is reached at

$$\kappa_{\theta\theta,1} \; = \; \kappa_{\theta\theta,max}, \tag{44}$$

which translates into (see eq. 40)

$$E_{crit} \; = \; (\frac{3}{4}\frac{U_1}{c})^2 \, E_{max} \; = \; E_{knee}. \tag{45}$$

We emphasize that two different basically geometric arguments lead to the same critical energy, $E_{knee}$, since we have derived the same critical energy from arguments about the polar cap. However, it is from this argument here, that the spectrum beyond the knee follows.

Above this critical particle energy then the particles average in their path over the small curvature scales, and perceive dominantly the full curvature of the system. Since at full curvature, $1/r$, there is no direction ambiguity, the same argument applied as above indicates that we have a net balance of $1 - 1/4$ for the curvature term, reducing it from its value (at $U_1/U_2 = 4$) of 2 to 3/4, i.e. in eq. (23) the numerical value of the first bracket goes from 3 to 1.75, thus reducing the value of $x$ from 2.25 to 1.729. This then leads to an overall spectrum of $E^{-2.735}$, before taking leakage into account. We emphasize that this result is based on putting all the parameters which go into it at their most simple or most extreme limit. Again, using the correction implied by a finite wind speed, of $V_W/U_1 = 0.2$ gives a modified spectrum of $E^{-2.866}$. Therefore we associate an error range of 0.13 with the spectral index derived above. We note in addition that in our model where the supernova explosions into the interstellar medium and the supernova explosions into stellar winds both contribute to moderate particle energies, these differences in the sources may imply differences in the propagation, which can only be quantified after a thorough investigation of the cosmic ray propagation in the Galaxy. Thus, finally, the spectrum is

$$E^{-3.07-0.07\pm0.07}, \tag{46}$$

with leakage accounted for. This is what we wanted to derive. As noted above the use of limiting arguments to derive the spectrum on either side of the knee implies that the knee itself may be quite soft, and thus curvature is to be expected (see Biermann 1994d).

## F. Assumptions and Systematic Uncertainties

The assumptions adopted are inspired by Prandtls mixing length approach; all use the key proposition that the **smallest dominant scale**, either in geometric length, or in velocity space, gives the natural transport coefficient. In this sense the assumptions are *derived from a basic principle* which we postulate.

Our basic **argument 1**, *based on observational evidence as well as theoretical reasoning*, is that for a cosmic ray mediated shock the *convective random walk* of energetic particles perpendicular to the *unperturbed* magnetic field can be described by a diffusive process with a downstream diffusion coefficient $\kappa_{rr,2}$ which is given by the thickness of the shocked layer and the velocity difference across the shock, and is independent of energy.

The upstream diffusion coefficient can be derived in a similar way, but with a larger scale based on the *same column density as in the downstream layer*. This leads to the second critical **argument 2**, namely that the upstream length scale is just $U_1/U_2$ times larger.

It must be remembered that the large intensity of convective turbulence increases the curvature: The characteristic scale of the turbulence is $r/4$ for strong shocks, again, as an example, in the case of the wind-SN, and thus the curvature is $4/r$ maximum. Half the maximum of the curvature allows for the net balance of gains and losses for the energy gain due to drifts (**argument 3**), and so we obtain then for the curvature $2/r$ which is twice the curvature without any turbulence; this increases the curvature term for the spectral range below the knee.

The diffusion tensor component $\kappa_{\theta\theta}$ can be derived similar to our heuristic derivation of the radial diffusion term $\kappa_{rr}$, again by using the smallest dominant scales. The characteristic velocity of particles in $\theta$ is given by the erratic part of the drifting, corresponding to spatial elements of different magnetic field direction. This is on average the value of the drift velocity $\mid V_{d,\theta} \mid$, possibly modified by the locally increased values of the magnetic field strength. The characteristic length is the distance to the symmetry axis $r \sin\theta$ (**argument 4**); this is the smallest dominant scale as soon as the thickness of the shocked layer is larger than the distance to the symmetry axis, *i.e.* $\sin\theta < U_2/U_1$.

Rapid convection also gives a competing diffusion in the $\theta$-direction, independent of particle energy; this will begin to dominate as soon as the energy dependent $\theta$-diffusion coefficient reaches this maximum at a critical energy. As long as the $\theta$-diffusion coefficient is smaller, it will dominate particle transport in $\theta$ and the upper limit derived here is irrelevant. When the $\theta$-diffusion coefficient reaches and passes this maximum given by the fast convection, then the particle in its drift will no longer see an increased curvature due to the convective turbulence due to averaging. The part of drift acceleration due to increased curvature is eliminated. Again, a detailed consideration of gains and losses of the drift energy gains leads to the spectrum of particles beyond the knee. The critical energy derived in this way is the same as that derived from a phase-space argument near the poles.

All these arguments are inspired by Prandtl's mixing length approach; all use the key proposition that the **smallest dominant scale**, either in geometric length, or in velocity space, gives the diffusive transport discussed. We assume this to be true even for the anisotropic transport parallel and perpendicular to the shock.

In addition, we i) use the simplified notion of a purely spherical shock; ii) ignore the modifications of the shock introduced by the cosmic rays themselves, except in the conceptual derivation of the initial argument, where the cosmic rays are critical for the instability; iii) use a test particle approach; iv) have not checked with a full 3D calculation on a supercomputer system that our conceptualization of the local physics actually yields what we argue; and v) also have not checked with such a calculation that our concept of the *smallest dominant scale* is as general as we use it. We note, however, that this concept is already contained in early descriptions of Prandtl's mixing length scale argument.

We have to emphasize very strongly that these uncertainties mean that the spectral indices derived for the powerlaw section of the various components of the cosmic rays correspond to a limiting argument: If things were really not as simple - and they are likely to be much more complicated - then the spectrum derived and any comparison with data has to be taken with considerable caution.

## G. Summary of the predictions

The proposal is that three sites of origin account for the cosmic rays observed, i) supernova explosions into the interstellar medium, ISM-SN, ii) supernova explosions into the stellar wind of the predecessor star, wind-SN, and iii) radio galaxy hot spots. Here the cosmic rays attributed to supernova-shocks in stellar winds, wind-SN, produce an important contribution at all energies up to $3 \; 10^9$ GeV.

Particle energies go up to 100 Z TeV for ISM-SN, and to 100 Z PeV with a bend at 600 Z TeV for wind-SN. Radiogalaxy hot spots go up to near or slightly beyond 100 EeV at the source. These numerical values are estimates with uncertainties of surely larger than a factor of 2, since they derive from an estimated strength of the magnetic field, and estimated values of the effective shock velocity (see above).

The spectra are predicted to be $E^{-2.75 \pm 0.04}$ for ISM-SN (paper CR

III), and for wind-SN $E^{-2.67-0.02\pm0.02}$ and $E^{-3.07-0.07\pm0.07}$ below and above the knee, respectively, and $E^{-2.0}$ at injection for radiogalaxy hot spots. The polar cap of the wind-SN contributes an $E^{-2.33}$ component (allowing for leakage from the Galaxy), which, however, contributes significantly only near and below the knee, if at all. The uncertainty in the radiogalaxy spectra will be discussed elsewhere. These spectra are for nuclei and are corrected for leakage from the galaxy. Electron spectra are discussed below.

The chemical abundances are near normal for the injection from ISM-SN, and are strongly enriched for the contributions from wind-SN. Helium and heavier elements are dominantly from wind-SN already at GeV particle energies. At the knee the spectrum bends downwards at a given rigidity, and so the heavier elements bend downwards at higher energy per particle. Thus beyond the knee the heavy elements dominate all the way to the switchover to the extragalactic component, which is, once again, mostly Hydrogen and Helium, corresponding to what is expected to contribute from the interstellar medium of a radiogalaxy, as well as from any intergalactic contribution mixed in (Biermann 1993b). This continuous mix in the chemical composition at the knee already renders the overall knee feature in a spectrum in energy per particle unavoidably quite smooth, a tendency which can only partially be off-set by the possible polar cap contribution, since that component also is strongest at a given rigidity.

We note that further uncertainties of the spectrum derive from a) the time evolution of any acceleration process as the shock races outward, b) the match between ISM-SN and wind-SN, c) the mixing of different stellar sources with possibly different magnetic properties, and d) the differences in propagation in any model which uses different source populations. These uncertainties translate into a distribution of powerlaw indices of the spectra, to curvature of the spectra, to a smearing of the knee feature, and to a smoothing of the cutoffs. Obviously, this is in addition to the underlying uncertainty associated with the concept of the *smallest dominant scale* itself.

There is one important prediction for stars: The particle energies of cosmic rays in the wind-component go up to 100 Z PeV, which *necessarily* implies that those massive stars which have strong winds into which they explode as supernovae, have appreciable magnetic fields at that stage: The magnetic field is required to be of order 3 Gauss at a distance of $10^{14}$ cm from the hydrostatic surface of the star; if the magnetic field topology is of a Parker type, *i.e.* is mostly tangential and depends on radius $r$ as $r^{-1}$, right down to near the surface of the star, as argued in paper CR II and below, then the surface magnetic fields are still quite moderate, *e.g.* of order a few thousand Gauss for Wolf Rayet stars. We will explore some of the consequences in the following.

## IV. Winds of massive stars

In this section we discuss the winds of massive stars, and the consequences of the proposition that the magnetic fields may not be totally negligible, *i.e.* may be of the order of 3 Gauss at $10^{14}$ cm radial distance in the stellar wind.

### A. Dynamo

First we propose that high magnetic fields can be generated inside massive stars so that later, when these interiors get exposed to become the surface of Wolf Rayet stars these high magnetic fields can help drive the wind. The dynamo mechanism is believed to act in the turbulent zone in the interior of rotating massive stars and, given a seed field such as may be produced by a battery mechanism in a rotating star, increases the magnetic field strength up to a maximum, which can be estimated in different ways. One possible limit is the dynamic pressure of the convective motions. The analogy with the interstellar medium suggests the limit where the Coriolis force equals the magnetic stresses (Ruzmaikin *et al.* 1988, Gilbert & Childress 1990, Gilbert 1991). The strength of the magnetic field can then be estimated from the condition, that the magnetic torque is limited by the Coriolis forces over the size of the convective region. This condition can be written as

$$B = (\Omega \, \rho_c \, R_{cc} \, v_t \, f_1)^{1/2}, \qquad (47)$$

where $\Omega$ is the rotation rate of the star, $\rho_c$ is the average density of the convection zone, $R_{cc}$ is the radius of the convective core, $v_t$ is the characteristic turbulent velocity, and $f_1$ is a correction factor of order unity in order to allow for a) structural variations ignored here, and also for b) the fact that the dominant length scale is likely to be smaller than the radius of the convective region. The turbulent velocity is estimated from the condition that turbulent convection transports all the luminosity $L$:

$$L = 4\pi \, R_{cc}^2 \, \rho_c \, v_t^3 \, f_2, \qquad (48)$$

where $f_2$ is also a correction factor of order unity to allow for structural variations; we use the average density and the radius of the entire convective region.

Models of Langer (1992, pers.comm.) provide the input together with data from the textbook of Cox & Giuli (1968):

$$R = 9.0 \, R_\odot \, (\frac{M_\star}{40 \, M_\odot})^{0.512}, \qquad (49)$$

$$M_{cc}/M = 0.628 \, (\frac{M_\star}{40 \, M_\odot})^{0.466}, \qquad (50)$$

$$R_{cc}/R = 0.384\,(\frac{M_\star}{40\,M_\odot})^{0.377}, \tag{51}$$

where $R$ and $M_\star$ are the stellar radius and mass, respectively. This leads to a magnetic field of

$$B = 1.9\,10^6 \alpha_{rot}^{1/4}\,f_1^{1/2}\,f_2^{-1/6}\,(\frac{M_\star}{40\,M_\odot})^{-0.058}\ \text{Gauss}, \tag{52}$$

where $\alpha_{rot}$ is the fraction of critical rotation at the surface, assumed to be solid body rotation. The same argument for Wolf Rayet stars (using data of Langer 1989, with Helium mass fraction $Y = 1$.) gives

$$B = 2.3\,10^7 \alpha_{rot}^{1/4}\,f_1^{1/2}\,f_2^{-1/6}\,(\frac{M_W}{5\,M_\odot})^{0.035}\ \text{Gauss}, \tag{53}$$

Hence this readily produces magnetic fields, dependent on the rotation rate of the star, of up to $2\,10^6$ Gauss for O stars, and about 10 times more for Wolf Rayet stars. In both cases the induced magnetic field is nearly independent of stellar mass. Critical rotation here means that the convective core has to rotate at an *angular* velocity which would correspond at the surface to critical rotation there; but in fact, as we will see below, it is not required that the actual surface rotates this fast.

In conclusion we find that we rather easily generate magnetic fields at levels deep inside the star beyond the local virial limit at the surface of the star, which gives of the order of a few times $10^4$ Gauss nearly independent of the rotation rate (Maheswaran & Cassinelli 1988, 1992). Rotationally induced circulations may be able to carry these magnetic fields to the surface of the stars on a time scale much shorter than the main sequence life time. In this transport the magnetic field is weakened by flux conservation and so surface fields in the range $10^3$ to $10^4$ Gauss are quite plausible, and are below the local virial theorem limit. Such values are all we require in the following. It is these weaker magnetic field strengths, which we will use in the following, so that the uncertainty of what really limits the dynamo mechanism, can be checked more from observation than theoretical faith. These estimates are valid for massive stars and extend clearly to below the mass range where we have Wolf Rayet stars as an important final phase of evolution. We thus expect many of the arguments in the following to hold generally for all massive single stars with extended winds, whether slow or fast, whether red or blue supergiant preceding the supernova explosion.

There is a further consequence for the generation of magnetic fields in white dwarfs: The most massive stars that do not become supernovae, but white dwarfs, are sufficiently massive to contain also convective cores which get exposed when the white dwarf is formed. Thus there ought to be a correlation between massive white dwarfs and the detection of strong magnetic fields; this is consistent with the observational data (Liebert 1992, pers.comm., Schmidt *et al.* 1992).

An interesting additional observation is the high incidence of kilo-Gauss fields on Am stars (Lanz & Mathys 1993); three of four Am stars searched have such strong fields, and there is evidence that the field is fairly disordered. The important point is that such stars have an outer radiative zone, and a core convection zone just like the more massive stars. Ultimately we may find a connection in the build-up of core magnetic fields in all upper main sequence stars with an inner convection zone.

It is not clear at present whether the magnetic fields on the surface of OB stars are fossil remnants from a pre-main sequence phase, or are fields that have been carried through the radiative zone by circulations. Calculations for this process similar to those of Charbonneau & MacGregor (1992) remain to be done and are required to demonstrate that this latter process is really possible. Such a transport mechanism has to remain an assumption at this stage.

## B. Wind-driving

In the standard version of the fast magnetic rotator theory (Hartmann & MacGregor 1982; Lamers & Cassinelli 1994, chapter 8) the magnetic field is calculated self-consistently and turns out to be radial close to the stellar surface and then starts bending at the critical point where the radial Alfvén velocity, the local corotation velocity and the wind velocity all coincide. This produces a long lever arm for the loss of angular momentum, and is the essence of the criticism of Nerney & Suess (1987) against the model. The spindown of fast magnetic rotator Wolf-Rayet stars is a topic addressed by Poe, Friend & Cassinelli (1989), where it is assumed that the field is radial close to the stellar surface.

However, the magnetically driven winds do not require the magnetic field to be radial near the surface. In a convection zone near the surface, it may be plausible to assume that the magnetic field is nearly isotropic in its turbulent character below the region where the wind gets started, and radial in the wind zone near to the star, leaving the radial magnetic field lines dominant for as long as the flow velocity is below the radial Alfvén speed. We note that the surface magnetic field structure on the Sun is mostly radial at the level where we can observe it, and we can only infer its structure below the surface from detailed modelling.

However, in a star, where the outer layers are radiative, it is not clear at all that the magnetic field is initially radial. Even the slightest differential rotation will tend to make a magnetic field, which originates in the central convective region, to be tangential. Even the acceleration into a wind does not obviously overturn this tendency completely; we just do not know. In fact, the recent model calculations of Wolf Rayet star winds by Kato & Iben (1992) suggest that the critical point of the wind where the wind becomes supersonic may be already inside the photosphere, and so it is reasonable to suppose that the other critical point where the wind speed exceeds the local radial Alfvén velocity may also be inside the star, if there is such a point at all - the radial flow velocity may be faster than the radial Alfvén velocity throughout. Similarly, the arguments by Lucy & Abbott (1993) which discuss the effect of multiple scattering in radiation driving, also lead to an effective initial acceleration of the wind, which may lead to a fast wind already near the photosphere.

In the following we propose to discuss such a wind, generalizing from Parker (1958) and Weber & Davis (1967). As in Weber & Davis we limit ourselves here to the equatorial region. We use magneto-hydrodynamics and Maxwells equations and thus have for the angular momentum transport $L_J$ per unit mass:

$$L_J = r\, v_\phi - (\frac{B_r}{4\,\pi\,\rho\,v_r})\, r\, B_\phi = const \,.$$  (54)

Here the components of the magnetic field are $B_\phi$ and $B_r$, and the components of the wind velocity are $v_\phi$ and $v_r$, while $\rho$ is the density and the index $a$ refers to a reference radius, which may or may not correspond to the stellar surface.

Introducing for the magnetic flux

$$F_B = r^2\, B_r = r_a^2\, B_{ra} = const,$$  (55)

with flux freezing

$$r\,(v_r B_\phi - v_\phi B_r) = -\Omega\, F_B = const \,,$$  (56)

and mass flux

$$\dot{M} = 4\,\pi\,\rho\,r^2\,v_r = 4\,\pi\,\rho_a\,r_a^2\,v_{ra} = const,$$  (57)

the tangential velocity can be written as

$$v_\phi = \frac{L_J}{r}\,(1 - \frac{F_B^2}{\dot{M}}\,\frac{\Omega}{L_J\,v_r})/(1 - \frac{F_B^2}{\dot{M}}\,\frac{1}{r^2\,v_r}).$$  (58)

The angular momentum loss can be written as

$$L_J = \epsilon\,\Omega\,r_a^2,$$  (59)

where $\epsilon$ defines the location from where the angular momentum loss is actually occurring. If there is no Alfvén critical point outside the star, then obviously

$$\epsilon \ < \ 1. \tag{60}$$

We note that the term $F_B^2/(\dot{M} v_r r^2)$ appearing in the expression for the tangential velocity can be rewritten with the surface radial Alfvén Mach number

$$M_{Ara} \ = \ v_{ra}/v_{Ara}, \tag{61}$$

where

$$v_{Ar} \ = \ B_r/(4\,\pi\,\rho)^{1/2}, \tag{62}$$

as

$$\frac{F_B^2}{\dot{M}} \frac{1}{v_r r^2} \ = \ \frac{1}{M_{Ara}^2} \frac{v_{ra}}{v_r} \frac{r_a^2}{r^2} \ = \ \frac{1}{M_{Ar}^2}. \tag{63}$$

Similarly we have the relationship

$$\frac{F_B^2}{\dot{M}} \frac{\Omega}{L_J v_r} \ = \ \frac{1}{M_{Ar}^2} \frac{r^2}{r_a^2} \frac{1}{\epsilon} \ = \ \frac{1}{M_{Ara}^2} \frac{v_{ra}}{v_r} \frac{1}{\epsilon}. \tag{64}$$

Thus the tangential velocity can be written as

$$v_\phi \ = \ \frac{L_J}{r} \left(1 - \frac{v_{ra}}{M_{Ara}^2 \, \epsilon \, v_r}\right)/(1 - \frac{1}{M_{Ar}^2}). \tag{65}$$

We can reasonably assume that the radial velocity is steadily increasing with radius. The tangential velocity should neither be negative nor exceed the rotational velocity of the star itself, and so we derive the conditions

$$M_{Ara} \ > \ 1, \tag{66}$$

and

$$M_{Ara}^{-2} \ < \ \epsilon \ < \ 1 \tag{67}$$

for the conditions envisaged here, that there is no Alfvén critical point outside the star. These conditions translate into

$$\frac{1}{M_{Ara}^2} \frac{v_{ra}}{v_r} \frac{1}{\epsilon} < 1, \tag{68}$$

and

$$\frac{1}{M_{Ar}^2} \ < \ 1. \tag{69}$$

It follows, $e.g.$, that $M_{Ar} \sim r$ asymptotically.

We define $U_\epsilon$ and $U_M$ and obtain

$$0 < U_\epsilon = 1 - \frac{\epsilon\, r_a^2}{r^2} < 1, \tag{70}$$

and

$$0 < U_M = 1 - \frac{1}{M_{Ar}^2} < 1. \tag{71}$$

The tangential magnetic field can then be written as

$$B_\phi = -\frac{F_B\, \Omega}{r\, v_r}\, (1 - \frac{\epsilon\, r_a^2}{r^2})/(1 - \frac{1}{M_{Ar}^2}), \tag{72}$$

and is thus also without change of sign outside the star. It is easy to verify that no magnetic flux is transported to infinity (see Parker 1958).

The radial momentum equation is

$$\begin{aligned}
(v_r \frac{d}{dr} v_r)\,(1 - \zeta \frac{c_s^2}{v_r^2}) &= \\
2\,\zeta \frac{c_s^2}{r} - \frac{G\, M_\star}{r^2} &+ \frac{F_{rad}\sigma_T N}{m_p c} + \\
\frac{v_\phi^2}{r} - \frac{1}{8\,\pi\,\rho\, r^2} &\frac{d}{dr}(rB_\phi)^2.
\end{aligned} \tag{73}$$

where $\zeta$ indicates the non-adiabacy of the flow:

$$\zeta = (\frac{dP}{d\rho})_{\text{along}\, r}/(\frac{dP}{d\rho})_{\text{adiabatic}}.$$

We will adopt $\zeta = 1$ for simplicity in the following.

In this equation the radiation force (flux $F_{rad}$) on both lines and continuum is given with the correction factor $N$ over the Thompson cross-section, also including the effect due to the chemical element composition being different from pure hydrogen; this factor $N$ depends on the physical state of the gas. The adiabatic speed of sound is $c_s$. We here proceed to evaluate the last term with the expressions already derived and discuss it in the context of the momentum equation.

The gradient term of $(r\, B_\phi)^2$ in the momentum equation can then be rewritten as (on the right hand side of the momentum equation)

$$\begin{aligned}
-\frac{1}{8\,\pi\,\rho\, r^2} &\frac{d}{dr}(rB_\phi)^2 = \\
(v_r \frac{d}{dr} v_r) \frac{1}{M_{Ara}^2} &(\frac{r_a \Omega}{v_{ra}})^2\, (\frac{v_{ra}}{v_r})^3 \frac{U_\epsilon^2}{U_M^3} \\
-2 \frac{1}{M_{Ara}^2}\, r_a \Omega^2\, (\frac{r_a}{r})^3 &\frac{v_{ra}}{v_r} \frac{U_\epsilon}{U_M^3}\, (\epsilon - \frac{1}{M_{Ara}^2} \frac{v_{ra}}{v_r}).
\end{aligned} \tag{74}$$

A second term on the right hand side of the momentum equation is the centrifugal force, which can be written as

$$\frac{v_\phi^2}{r} = r_a \Omega^2 \left(\frac{r_a}{r}\right)^3 \frac{1}{U_M^2}\left(\epsilon - \frac{1}{M_{Ara}^2}\frac{v_{ra}}{v_r}\right)^2. \tag{75}$$

The last term in brackets never goes through zero outside the star because of the conditions we have set above for $\epsilon$ and $M_{Ara}$. Comparing now the centrifugal term with the second term from the gradient of the tangential magnetic field, we note that for $dv_r/dr$ approaching 0 these two terms differ by the factor

$$-\frac{1}{2}\frac{U_M}{U_\epsilon}\left(\epsilon\, M_{Ara}^2\frac{v_r}{v_{ra}} - 1\right), \tag{76}$$

where $\epsilon M_{Ara}^2 > 1$. If $v_r/v_{ra}$ exceeds a value of 3 at large radius $r$, then the centrifugal term, which provides some acceleration, dominates over the other term at large $r$. Here we have ignored all the terms of order unity that approach unity with both the radius $r$ and the radial velocity $v_r$ becoming large. Even close to the star the centrifugal force may dominate.

The first term in the gradient of the tangential magnetic field is more interesting, however. This term becomes a summand to the various terms multiplying $(v_r\frac{d}{dr}v_r)$ and has there, on the left hand side, a minus sign and is to be compared with unity in the case that we are already at supersonic speed. This may be attained through radiation driving with multiple scattering.

We define an Alfvén Machnumber with respect to the total magnetic field with

$$M_A^2 = \frac{v_r^2 4\pi\rho}{B_\phi^2 + B_r^2}. \tag{77}$$

Generally we have obviously

$$M_A < M_{Ar}. \tag{78}$$

With this generalized Alfvén Machnumber we can rewrite the entire factor to $(v_r\frac{d}{dr}v_r)$ in a simple form

$$1 - \frac{\zeta}{M_s^2} - \left(\frac{M_{Ar}^2}{M_A^2} - 1\right)/(M_{Ar}^2 - 1), \tag{79}$$

where $M_s$ is the sonic Mach number for the radial flow velocity. Critical points appear whenever this expression goes through zero or a singularity. This expression is easily seen to be equivalent to eq. (6) of Hartmann & MacGregor (1982); they, however, find a magnetic field which is initially radial and neglected radiative forces. This also shows that we have the critical point of Weber & Davis (1967) again for $\frac{\zeta}{M_s^2} \ll 1$, when and if $M_{Ar} = 1$, but we have in our case another critical point at $M_A = 1$, which is the fast magnetosonic point (see Hartmann & MacGregor 1982). In order to make a realistic judgement on this question, we would have to combine the model of Kato & Iben (1992) or a detailed calculation of multiple scattering (Lucy & Abbott 1993) to drive winds initially with a realistic treatment of the magnetic field inside the star, so that we can derive the range of possible properties of the magnetic field near the surface of the star.

However, there are a few general conclusions one can draw already: If the magnetic field inside the star begins as a dominantly tangential field, then it is by no means clear whether there is any point at which we have $M_{Ar} < 1$, either inside or outside the star; going outwards the unwinding of the magnetic field is intimately coupled to the initially weak radial flow, and so $M_{Ar} > 1$ may hold throughout. In that case the term derived from the radial gradient of the tangential magnetic field goes through unity only when the generalized Alfvén Mach number goes through unity. If, as we argued, $M_{Ar} > 1$ possible throughout, then, again going outwards with radius, we start with a low generalized Alfvén Mach numbers $M_A$, either supersonic or subsonic flow, the entire expression is negative, matching the negative gravitational force on the right hand side. Far out, both sonic and generalized Alfvén Mach-numbers are large, and so the expression is positive. It follows that the expression has to go through zero. The condition $M_A = 1$ here is the fast magnetosonic point, and the radial velocity there corresponds to the Michel-velocity. Or, in other words, the radial velocity is equal to the total Alfvén velocity.

We know from observations that the winds in OB and Wolf Rayet stars are very strongly supersonic; using our model we deduce that the radial velocity is weakly super-Alfvénic with respect to the tangential (dominant) magnetic field component, and so we expect far outside the star a configuration where $M_s >> 1$, $M_{Ar} >> 1$ and $M_A \gtrsim 1$, but this latter condition may not necessarily be satisfied by much. This means that the expression is near to unity and somewhat smaller than unity. This entails the condition that the right hand side has to be positive, which is readily interpreted as possibly arising from line radiation, because that term is the only one which has the same radial dependence as the gravitational force. Comparing then the effect of line driving (Lucy & Solomon 1970, Castor *et al.* 1975) with and without such a magnetic field, the net effect is an amplification in the sense that for $M_{Ar}^2 \gg 1$ the velocity gradient is asymptotically increased by

$$1/(1 - \frac{1}{M_A^2}). \qquad (80)$$

For $M_A$ close to unity, this factor can be arbitrarily large, and thus illustrates the amplification possible from the pressure gradient of a tangential magnetic field. Obviously, such a large amplification is present only for a small radial region, and so the final wind velocity can still likely be not much more than the tangential Alfvén velocity (see below). We thus argue that there is a magnetic field configuration where the field is mostly tangential already inside the star, and remains mostly tangential outside the star, where the initial acceleration of the wind is done by the opacity mechanism discussed by Kato & Iben (1992) or by multiple scattering (Lucy & Abbott 1993) in a line-driving modus; outside the star we have a line-driven wind, but the amplification by the effect of the pressure gradient of the tangential magnetic field really produces the large momentum in the wind.

To see the properties of the equations better, we simplify by dropping all right hand terms in the differential equation except for the gravitational and the radiative force, and introduce characteristic length and velocity scales

$$r_\star = \left(\frac{GM_\star}{v_{ra}^2}\right)\left(\frac{v_{ra}}{v_{Ara}}\right)^{4/3}\left(\frac{v_{ra}}{r_a\Omega}\right)^{4/3}, \qquad (81)$$

and

$$v_\star = v_{ra}\left(\frac{v_{Ara}}{v_{ra}}\right)^{2/3}\left(\frac{r_a\Omega}{v_{ra}}\right)^{2/3}, \qquad (82)$$

which is easily recognized as the Michel velocity (Hartmann & Mac-Gregor 1982). This equation corrects a misprint in the corresponding expression in paper CR II, in that the second exponent is 2/3 instead of 1/3. These normalizations lead to the same overall dimensionless form of the momentum equation as in Owocki (1990), since we have to multiply the entire eq. (73) with

$$\frac{r^2}{r_\star v_\star^2} = \frac{r^2}{GM_\star}.$$

The Eddington luminosity is given by

$$L_{edd} = \frac{4\pi GM_\star m_p c}{\sigma_T}, \qquad (83)$$

where $\sigma_T$ is the Thompson cross section. In the approximation that $U_\epsilon = U_M \simeq 1$ and $r\Omega \gg v_r$ the Michel velocity corresponds to the overall Alfvén velocity.

Next we write down the whole differential wind equation in dimensionless form:

$$-\frac{1}{2}\frac{dy^2}{d(1/x)}[1 - \frac{\zeta}{M_s^2} - (\frac{M_{Ar}^2}{M_A^2} - 1)/(M_{Ar}^2 - 1)] =$$

$$+ 2\zeta c_{sa}^2 r_a (\frac{r_a}{r})^{1/3} (\frac{v_{ra}}{v_r})^{2/3} \frac{1}{GM_\star}$$

$$+ \frac{LN}{L_{edd}} - 1 \tag{84}$$

$$+ \Omega^2 r_a^3 (\frac{r_a}{r}) \frac{1}{U_M^2} (\epsilon - \frac{1}{M_{Ara}^2} \frac{v_{ra}}{v_r})$$

$$[-2\frac{U_\epsilon}{U_M}\frac{v_{ra}}{v_r}\frac{1}{M_{Ara}^2} + \epsilon - \frac{1}{M_{Ara}^2}\frac{v_{ra}}{v_r}]\frac{1}{GM_\star}$$

Here we have to note that in a full treatment of this equation the temperature enters in the speed of sound, here approximated by an adiabatic law with gas constant 5/3, the state of ionization enters into the factor $N$, which can be highly space dependent (see, *e.g.*, Lucy & Abbott 1993), and that hence we can discuss here only a very simplified form of this equation. We approximate this expression in several steps in order to finally write down only two limiting forms.

In the approximation that the sonic and the radial Alfvénic Mach-number are both large already on the surface, the left hand side simplifies to

$$-\frac{1}{2}\frac{dy^2}{d(1/x)}(1 - \frac{1}{M_A^2}) \tag{85}$$

showing only the fast magnetosonic point as a remaining singularity. At that point the right hand side has to go from negative to positive, which could happen in several ways. First of all, the pressure term is clearly unimportant, since we already neglected the subsonic regime. The last term, which derives from the centrifugal force and the second part of the magnetic field pressure gradient, is asymptotically positive; on the surface this term may be negative. Writing

$$\epsilon = \frac{E}{M_{Ara}^2} < 1, \tag{86}$$

with the condition

$$M_{Ara}^2 > E > 1, \tag{87}$$

we find the condition on the surface

$$E < \frac{3M_{Ara}^2 - 1}{M_{Ara}^2 + 1}. \tag{88}$$

for the last term in eq. (84) to be negative on the surface. It thus depends on a detailed consideration of the inner structure of the star and its rotation law to determine whether this term can be sufficiently large and negative on the surface to compensate the radiative driving term, and so to locate the fast magnetosonic point.

Another possibility is that the radial dependence of the radiative driving term, *i.e.* the radial variation of the factor $N$ determines the location of the fast magnetosonic point. In that case we may be allowed to neglect the centrifugal and the remaining term of the magnetic pressure gradient.

The differential wind equation then reads finally in dimensionless length $x$ and velocity $y$

$$\frac{1}{2}\left(1-\frac{1}{y^3}\right)\frac{dy^2}{d\left(1/x\right)} \;=\; -f_{rad}, \tag{89}$$

with

$$f_{rad} = \frac{NL}{L_{edd}} - 1. \tag{90}$$

We note that the radial dependence of the magnetic field strength leads asymptotically to

$$\left(\frac{M_{Ar}^2}{M_A^2}-1\right)/(M_{Ar}^2-1) = \left(y^3\,\frac{U_M^3}{U_c^2}\right)^{-1} \simeq y^{-3}, \tag{91}$$

in the limit, that $U_M^3/U_c^2 \to 1$.

Clearly we require then that the radiative force dominates over gravity, by however little, to obtain the correct sign of the right hand side, thus $f_{rad} > 0$. We assume here for illustration that $N$ is constant with $r$.

The analytic solution is

$$\frac{1}{2}\left(y^2-y_i^2\right)+\frac{1}{y}-\frac{1}{y_i} \;=\; -f_{rad}\left(\frac{1}{x}-\frac{1}{x_i}\right), \tag{92}$$

where the index $i$ refers to an initial state. For $y \gg 1 > y_i$ and $x \gg x_i$ the limiting solution is

$$y \;=\; \sqrt{2\left(\frac{1}{y_i}+f_{rad}\,\frac{1}{x_i}\right)}. \tag{93}$$

We note right away that both effects contribute to a faster wind. Thus, the analytic solution yields then two extreme possibilities, either

$$v_{r\infty} \;=\; v_e\left(\frac{NL}{L_{edd}}-1\right)^{1/2}, \tag{94}$$

or

$$v_{r\infty} = \sqrt{2} \left(\frac{v_{Ara}}{v_{ra}}\right)^{1/3} \left(\frac{r_a \Omega}{v_{ra}}\right)^{1/3} v_\star, \qquad (95)$$

where we have used the definition

$$v_e = \left(\frac{2GM_\star}{r_a}\right)^{1/2} \qquad (96)$$

for the surface escape velocity from the star. Obviously, we require that

$$2^{3/2} \left(\frac{v_{Ara}}{v_{ra}}\right) \left(\frac{r_a \Omega}{v_{ra}}\right) > 1, \qquad (97)$$

in order to have a final velocity above the Michel velocity. We note that

$$\frac{\Omega r}{v_r} = -\frac{B_\phi}{B_r} \frac{U_M}{U_\iota}, \qquad (98)$$

for any $r$ and also on the surface, and thus the condition is likely to be fulfilled for a magnetic field configuration which is highly tangential at the stellar surface. By the same condition, the initial surface velocity is below the Michel velocity (except for the factor 2 which is of order unity). This is necessary for the flow to have a transition through the fast magnetosonic point. This sharpens the approximations introduced above. We note that in the magnetic driving case, the final velocity far outside can also be written as

$$v_{r\infty} = v_e \sqrt{2} \frac{v_{Ara}}{v_{ra}} \frac{r_a \Omega}{v_e}, \qquad (99)$$

where both factors to $v_e \sqrt{2}$ are less than unity in our approximation, and so the case of pure magnetic driving in this approximation may yield lower velocities than radiation driving at large distances from the star in this very simple first approximation. The transition between the two extreme cases is given by

$$2 \left(\frac{v_{Ara}}{v_{ra}} \frac{r_a \Omega}{v_e}\right)^2 \simeq \frac{NL}{L_{edd}} - 1. \qquad (100)$$

We have thus in this approximation two possible extreme solutions: First, for small magnetic field, the Alfvén velocity drops out and the line driving is the regulating agency; second, for large magnetic field, the terminal wind velocity is not far from the tangential Alfvén speed, and the line driving effect is important but not dominant in that it provides the source of momentum to be amplified. This latter picture is the concept we propose to explain the momentum in the winds of Wolf Rayet stars. This is different from the solutions of Hartmann &

MacGregor (1982) in the sense, that in their wind solutions the main acceleration of the wind is between the slow magnetosonic point and the Alfvén point, both of which are not outside the star in the case which we discuss; our solutions are similar, however, in that both in their case as here, the terminal velocity of the wind in the case of interest is not far from the the total Alfvén velocity. In this picture the angular momentum loss of the star refers to a characteristic level inside the star and so the criticism of Nerney & Suess (1987) is countered; the angular momentum loss of the star through mass loss is reduced.

Introducing then the radiative driving in the form of Castor *et al.* (1975), and using the nomenclature of Owocki (1990), we have for the radiation driving the following form

$$\frac{N L}{L_{edd}} \; = \; C_{CAK} \, (-y \, \frac{dy}{d(1/x)})^\alpha + \Gamma, \tag{101}$$

where $C_{CAK}$ is related to the mass loss rate (Owocki 1990, there eq. 14) and $\Gamma = L/L_{edd}$ describes the effect of the continuum radiation, and thus in a second approximation

$$\frac{1}{2} \, (1 - \frac{1}{y^3}) \, \frac{dy^2}{d(1/x)} \; = \; -C_{CAK} \, (-y \, \frac{dy}{d(1/x)})^\alpha + 1 - \Gamma. \tag{102}$$

This replaces then eq. (13) of Owocki (1990). Here $C_{CAK} \, (-y \frac{dy}{d(1/x)})^\alpha$ describes the driving by radiative forces in the approximation by Castor *et al.* (1975), and describes the spatial dependence of $N$ above. This shows that we have a critical point where $y = 1$ and where $C_{CAK} \, (-y \frac{dy}{d(1/x)})^\alpha = 1 - \Gamma$. We have to ask here, whether we have a critical point at all, where $M_A = 1$. Consider the case of low magnetic field: Then the initial surface flow velocity may well be super-Alfvénic, and so $y_a > 1$, and if in fact $y_a \gg 1$, then the magnetic field can be neglected. However, if the initial velocity is sub-Alfvénic and the final radiation-driven wind is super-Alfvénic, then magnetic driving certainly cannot be neglected. These conditions can be written as

$$\frac{v_{ra}}{v_\star} \; = \; (\frac{v_{ra}}{v_{Ara}})^{2/3} (\frac{v_{ra}}{r_a \Omega})^{2/3} \; < \; 1, \tag{103}$$

and

$$\frac{v_{CAK}}{v_\star} \; = \; \frac{v_e}{v_{ra}} (\frac{v_{ra}}{v_{Ara}})^{2/3} (\frac{v_{ra}}{r_a \Omega})^{2/3} \sqrt{\frac{\alpha}{1-\alpha}} \; > \; 1. \tag{104}$$

If these two conditions are fulfilled, then we certainly have a critical point, where $y = 1$. Since $\Omega r_a/v_{ra}$ proportional to the ratio of the tangential over radial component of the magnetic field on the reference level of the star (see eq. 98), a sufficiently dominant tangential field will always ensure, that the first condition is fulfilled. Strong wind driving corresponds to $\alpha$ close to unity, and so in many cases the second condition will also be fulfilled. In the case that the first condition is fulfilled, but not the second, the effects of the magnetic field also cannot be neglected, but a full solution is required to discuss this at the required depth. A full solution of this equation (102) remains to be done.

Using the observed wind velocities then gives an estimate for the Alfvén velocity, and thus implies - assuming our wind driving theory to be correct - that the magnetic field is quite high, and of the order of 3 Gauss at the fiducial radius of $10^{14}$ cm. Or, to turn the argument around, the magnetic field strengths implied by our cosmic ray arguments provide a possible explanation for the origin of the momentum of Wolf Rayet star winds. On the surface of the star the magnetic field strength implied is of order a few thousand Gauss, quite easily within the limits implied by the surface virial theorem. Also the surface of the star may not rotate as fast, but the inside could still rotate at an angular velocity corresponding to near critical at the surface. We emphasize that here our proposed wind driving theory - if true - provides an independent argument on the magnetic field strength on the surface of Wolf Rayet stars: If it can be shown to be the correct physical interpretation, then these magnetic field strengths follow. On the other hand, if observations show such magnetic field strengths to be too high, then both the wind theory proposed here, as well some of the cosmic ray arguments may fail.

## C. Radioemission from stars

In this section we derive the basic expressions for the luminosity, spectrum, and time dependence of the nonthermal radio emission from single spherical shocks in the winds of single massive stars. In the subsequent sections we then use these expressions to discuss radiosupernovae, young radio supernova remnants in starburst galaxies, Wolf Rayet stars, and OB stars. The radioemission of the nova GK Per (Seaquist *et al.* 1989) remains to be discussed in the detail required to match it with what we know about cataclysmic variables, close binary star systems with an accretion disk around a white dwarf (Biermann, Strom, Falcke 1994).

Electrons suffer massive losses due to Synchrotron radiation and so they can achieve only energies up to that point where the acceleration and loss times become equal; this is a strong function of latitude since the magnetic field varies rather strongly with latitude. The shock transfers a fraction $\eta$ of the bulk flow energy into relativistic electrons. The emissivity of an electron population with the spectrum

$$N(\gamma)\, d\gamma \;=\; C\, \gamma^{-p}\, d\gamma \tag{105}$$

is given by (Rybicki & Lightman 1979) in cgs units for $p = 7/3$ by

$$\epsilon_\nu \;=\; 3.90\; 10^{-18}\, C\, B^{5/3}\, \nu^{-2/3}\;\mathrm{erg\,sec^{-1}\,cm^{-3}\,Hz^{-1}}. \tag{106}$$

The case of a wind speed equal to the shock velocity and the limit of a strong shock in a gas of adiabatic index $5/3$ leads to an optically thin synchrotron spectrum of $-0.735$, and thus $p = 2.47$. The corresponding expressions for $p = 2.47$ are given in paper CR II. Similarly the absorption coefficient for synchrotron self absorption is for $p = 7/3$

$$\kappa_{\nu, syn} \;=\; 1.23\; 10^{12}\, C\, B^{13/6}\, \nu^{-19/6}\,\mathrm{cm^{-1}}. \tag{107}$$

In all these expressions we have averaged the aspect angle. We will use here, for illustration, the limit of large shock speeds for the discussion of the radiosupernovae, where the radio spectral index is $-2/3$. The free-free opacity is given by

$$\kappa_{\nu, ff} \;=\; 0.21\, T_e^{-1.35}\, \nu^{-2.1}\, n_e^2\,\mathrm{cm^{-1}}, \tag{108}$$

in an approximation originally derived by Altenhoff $et\ al.$ (1960) and widely disseminated by Mezger & Henderson (1967). Here $T_e$ and $n_e$ are the electron temperature and density, and the approximation has been used that the ion and electron density are equal, and that the effective charge of the ions is $Z = 1$. We note that for cosmic abundances we have the following approximations (Schmutzler 1987) assuming full ionization: $\rho = 1.3621\, m_H\, n$, $n_e = 1.181\, n$, and $n_i = 1.086\, n$, where $n$ is the total Hydrogen density (particles of mass $m_H$ per cc), and $n_i$ the ion density. With these approximations for full ionization the free-free opacity would increase a factor of two over the approximation by Altenhoff $et\ al.$ (1960), which, however, is compensated by the incomplete ionization (see, again, the calculations by Schmutzler 1987) in the photon ionized regions (here, winds) near massive stars. In the following we will use the approximation by Altenhoff $et\ al.$ replacing $n_e = n$.

The synchrotron luminosity in the optically thin limit is then given by an integration over the emitting volume, which we take in the context of our simplified picture to be a spherical shell of thickness $r/4$. The radio emission only arises from that part of the shell where the shock velocity is larger than the local Alfvén velocity and where the synchrotron loss time is longer than the local acceleration time. These conditions can lead to a restricted range in latitude for the emission (see, $e.g.$, Nath & Biermann 1994b, Biermann, Strom, Falcke 1994).

The luminosity is then given by the integral over the latitude dependent emissivity. Here we have to normalize the cosmic ray electron density to the shock energy density, using our induced latitude dependence. We will concentrate on the case, when the energetic particle density is stronger near the equator, and discuss the opposite case briefly at the end. Using a lower bound for the electron spectrum much below the rest mass energy and an upper bound much above that value we have for the constant $C$ then the expression

$$C(\mu) = C_o \,(1-\mu^2)^{3/2} \tag{109}$$

with $\mu_\star \geq \mu \geq 0$ where $\mu_\star$ refers to that latitude where the latitude dependent acceleration breaks down:

$$(1-\mu_\star^2)^{1/2} = \frac{3}{4}\,\frac{U_1}{c} \ll 1\;, \tag{110}$$

from our argument about the knee energy (see section 8 of paper CR I) for protons and other nuclei. At the equator the energy density of the electrons can be written as

$$C_o = (p-2)\eta\,\rho U_1^2/(m_e c^2) \tag{111}$$

where we use just the relativistic part of the distribution function and incorporate the uncertainty on the existence and strength of any sub-relativistic part of the electron distribution function in the factor $\eta$. The integration over latitudes leads to an integral correction factor of $\frac{1}{2}B(\frac{p+11}{4},\frac{1}{2})$, where $B(z_1,z_2)$ is the Beta-function, and $p$ the powerlaw index of the electron distribution function. For the powerlaw index used here, $p = 7/3$, this integral is very close to 0.50 in value. We also have

$$\rho = \frac{\dot{M}}{4\pi r^2 V_W}. \tag{112}$$

Assuming that all latitudes contribute up to $\mu_\star \lesssim 1$, the final expression for the luminosity is then

$$L_\nu(nth) = 8.1\,10^{24}\ \mathrm{erg/sec/Hz}$$
$$\eta_{-1}\,\frac{\dot{M}_{-5}}{V_{W,-2}}\,U_{1,-2}^2\,B_{0.5}^{5/3}\,r_{14}^{-2/3}\,\nu_{9.7}^{-2/3}\;. \tag{113}$$

Here the mass loss is in units of $10^{-5}\ \mathrm{M_\odot/yr}$, the unperturbed magnetic field strength at the reference radius of $10^{14}$ cm in units of of 3 Gauss, the wind and the shock velocity in units of 0.01 $c$, and as reference frequency we use 5 GHz.

Here we note that this emission might become optically thick both to free-free absorption in the lower temperature region outside the shock, or to to Synchrotron self absorption inside the shocked region. In the approximation, that we use the equatorial region in the slab model (*i.e.*, direct central axis radial integration), we obtain for the critical radius $r_{1,ff}$, where the free-free absorption has optical thickness unity:

$$r_{1,ff} \; = \; 1.35 \, 10^{14} \, (\frac{\dot{M}_{-5}}{V_{W,-2}})^{2/3} \, \nu_{9.7}^{-0.70} \, \text{cm.} \tag{114}$$

We have used here the analytical approximation for the free-free opacity introduced above and assumed a temperature of $T_e = 2 \, 10^4$ K.

Similarly, for synchrotron self absorption the critical radius $r_{1,syn}$ is given for $p = 7/3$:

$$r_{1,syn} \; = 3.99 \, 10^{14} \, \text{cm}$$
$$\eta_{-1}^{0.316} \, (\frac{\dot{M}_{-5}}{V_{W,-2}})^{0.316} \, U_{1,-2}^{0.632} \, B_{0.5}^{0.684} \, \nu_{9.7}^{-1} \tag{115}$$

The maximum synchrotron luminosity is given by the dominant absorption process; which one is stronger is given by comparing the relevant radii; in the numerical example synchrotron self absorption happens to be stronger. Free-free absorption is dominant for $p = 7/3$ if

$$\frac{\dot{M}_{-5}}{V_{W,-2}} \; > \; 22.0 \, \eta_{-1}^{0.901} \, U_{1,-2}^{1.802} \, B_{0.5}^{1.951} \, \nu_{9.7}^{-0.856}. \tag{116}$$

We note that the parameter $\eta$ might well be quite low.

In the following we give then the maximum luminosities calculated by using the proper optical depth for a central axis approximation in a slab geometry for the radiative transfer at the equator; this gives an additional factor of about 2/3 (see below). However, for the total emission we do integrate properly over all latitudes, while for the absorption we approximate by using the equatorial values. This is a fair approximation, since both emission and absorption decrease towards the poles, but the absorption decreases even faster. The exact numerical value in our approximation is used here.

In the case that free-free absorption dominates, the maximum luminosity is then given for $p = 7/3$ by

$$L_\nu(nth) \; = 3.8 \, 10^{24} \, \text{erg} \sec^{-1} \text{Hz}^{-1}$$
$$\eta_{-1} \, (\frac{\dot{M}_{-5}}{V_{W,-2}})^{0.556} \, U_{1,-2}^2 \, B_{0.5}^{1.667} \, \nu_{9.7}^{-0.200} \tag{117}$$

In the case that Synchrotron self absorption dominates, these maximum luminosities are given by

$$L_\nu(nth) = 2.2\ 10^{24}\,\mathrm{erg\,sec^{-1}\,Hz^{-1}}$$

$$\eta_{-1}^{0.789}\,(\frac{\dot{M}_{-5}}{V_{W,-2}})^{0.789}\,U_{1,-2}^{1.579}\,B_{0.5}^{1.211} \tag{118}$$

Here we do not consider mixed cases.

Obviously, the ratio of mass loss rate and wind velocity can be different by many orders of magnitude among predecessor supernova stars, and their numerical values will have to be argued on the basis of observations.

It is useful to also calculate the thermal radio emission, using the same standard parameters, then adjust our parameters to a realistic range, and then compare the nonthermal luminosities. The thermal emission can be properly integrated, allowing for the sphericity of the wind structure (Biermann *et al.* 1990) to give:

$$L_\nu(th) = 4\pi^2\,B_\nu(T)\,r_{1,ff}^2\,\Gamma(\frac{1}{3}) \tag{119}$$

With our standard parameters this luminosity in a steady wind,

$$L_\nu(th) = 3.0\ 10^{17}\,(\frac{\dot{M}_{-5}}{V_{W,-2}})^{4/3}\,\nu_{9.7}^{+0.60}\ \mathrm{erg/sec/Hz}\ , \tag{120}$$

is weakly dependent on electron temperature.

In the following we make a number of consistency checks on the basic notions used above:

First, we note that shocks can only exist when the shock velocity is larger than the Alfvén velocity. Since in our wind driving theory developed above the wind velocity is just a little larger than the Alfvén velocity itself, this implies that the shock velocity in the frame of the flow has to be at least the same velocity as the wind, and so - within our analytical approximations - we have the condition that always

$$U_1 \geq V_W\,, \tag{121}$$

where the limit of the equality corresponds to a spectral index for the Synchrotron emission of $-0.735$ and in the limit of large shock velocity to $-2/3$. Here we have to note that the Alfvén velocity is colatitude $\theta$ dependent and is proportional to $sin\,\theta$. Thus, even for lower shock speeds, there is a latitude range where a shock can be formed, but then the luminosity is very much reduced.

Second, we have to check that the Synchrotron loss time is larger than the acceleration time, because otherwise we would not have any electrons at the appropiate relativistic energies. This condition can be rewritten as

$$\nu_{9.7} B_{0.5}^3 \lesssim U_{1,-2}^2 r_{14}. \qquad (122)$$

This is the condition at the equator, and the condition gets weaker at higher latitudes. This makes it obvious, that for reasonable ranges of the parameters acceleration can succeed to the required electron energies. It also shows that at higher frequencies or smaller radii the condition would fail. At smaller radii the emission is usually optically thick (see above), and higher frequencies are as yet difficult to observe at the required sensitivity. The implied high frequency cutoff in the observed synchrotron spectra would, however, be an important clue.

Third, we have to check whether the implication that the shock speeds are typically similar to the wind speeds, is supported by data on Wolf Rayet stars, and their theoretical understanding. The wind calculations with shocks suggest that the typical shock velocities are indeed of order the wind speed itself or somewhat higher (Owocki *et al.* 1988), and so we expect a range in radio spectral indices of $-0.667$ to $-0.735$ or steeper if the shocks are not strong, *i.e.* if $U_1/U_2 < 4$. We note that the actual value of the nonthermal luminosity is changed only moderately in this range of spectral indices.

Fourth, we have to discuss optical thickness effects in more detail: A shock travels from the region inside of where free-free absorption dominates through this region to the outside. The emission then is first weak and has a steep spectrum due to the strong frequency dependence of the free-free absorption and the exponential cutoff induced, then approaches a peak in emission with a spectral index approaching $-0.67$ to near about $-0.735$ or steeper as discussed above, and then becomes weaker with the optically thin spectrum. At the location where our ray encounters the shock, the temperature of the gas increases drastically, and so the differential optical depth for free-free absorption goes to zero. Hence we have the simple case that we have emission inside the shock and absorption outside. We limit ourselves to the central axis as a first approximation, and also neglect the spatial variation of the emission itself (see above). This is equivalent to pure screen-like absorption, however with a screen which extends from the shock to the outside and so is variable. The radio luminosity is given by (using free-free absorption)

$$L_\nu(nth) = e^{-\tau_\nu} L_\nu(nth, no\,abs). \qquad (123)$$

We have the spectral index $-\alpha_{thin}$ in the optically thin regime, and the time dependence of the spectral index given by

$$\alpha(\nu) = -\alpha_{thin} (1 - (t_\nu^*/t)^3). \qquad (124)$$

The spectral index is positive and very steep at first and then goes through zero to become slowly negative approaching asymptotically the optically thin spectral index. The nonthermal luminosity considered as a function of time sharply rises at first, then peaks at optical depth $\alpha_{thin}/3$, obviously strongly dependent on frequency, and finally drops off as $1/t^{\alpha_{thin}}$:

$$\frac{d\ln L_\nu(nth)}{d\ln t} = -\alpha_{thin}\left(1 - \frac{10}{7}(t_\nu^*/t)^3\right). \tag{125}$$

Here we see that the time dependence of the flux density at a given frequency and the time dependence of the spectral index are closely related. The time of luminosity maximum depends on frequency as

$$t_{max\,L} \sim \nu^{-0.7} \tag{126}$$

from the frequency dependence of the optical depth

$$\tau_\nu \sim t^{-3}\,\nu^{-2.1}. \tag{127}$$

These relationships can be used to check on the importance of free-free absorption in our approximations.

Synchrotron self absorption is due to internal absorption inside the shell, and so again in the central axis approximation we have

$$L_\nu(nth) = \frac{1 - e^{-\tau_\nu}}{\tau_\nu}\,L_\nu(nth,\,no\,abs). \tag{128}$$

This then results in the time dependence for the spectral index of

$$\alpha(\nu) = \frac{5}{2} + \frac{5 + 2\alpha_{thin}}{2}\,\tau_\nu - \frac{5 + 2\alpha_{thin}}{2}\,\frac{\tau_\nu}{1 - e^{-\tau_\nu}}, \tag{129}$$

with

$$\tau_\nu \sim t^{-(5+2\alpha)/2}\,\nu^{-(5+2\alpha)/2}. \tag{130}$$

The time dependence of the double logarithmic derivative of luminosity on time is exactly the same

$$\frac{d\ln L_\nu(nth)}{d\ln t} = \alpha(\nu). \tag{131}$$

It follows that the time of maximum depends on frequency as

$$t_{max\,L} \sim \nu^{-1}. \tag{132}$$

This is a characteristic feature for synchrotron self absorption in the approximation used (and well known from radioquasars) and differs from the case considered above, for free-free absorption.

Thus, the true maximum of the luminosity is given by an optical depth less than unity, which gives a correction factor to the luminosities introduced above (eqs. 117 and 118) of about 2/3; we have corrected the luminosities introduced there for this factor.

Finally, we have to comment on the sign of the magnetic field. We have used here throughout the assumption that the magnetic field is oriented such that the drifts are towards the equator for the particles considered. Clearly, since we consider stars with a magnetic field driven by turbulent convection in a rotating system, we can expect that there are sign reversals of the magnetic field just as on the Sun. For the other sign of the magnetic field and the other drift direction there is by many powers of ten less nonthermal emission and so it is to be expected that at any given time we should detect at most half of the stars in nonthermal radio emission; occasionally we might even catch a shock travelling through the region in the wind where the sign is reversing itself, because the wind mirrors the time history of the star in terms of magnetic field. In the case, that the shock travels through a layer bounded by magnetic field reversals both inside and outside, then it becomes important to ask what the time scale of particle acceleration is relative to the time scale of traversing this layer; particle acceleration at high energies may be severely limited, if such layers are thin.

In the following we will first discuss the radio observations of Wolf Rayet stars, then OB stars, and finally supernovae.

The theoretical luminosities derived can be compared with the observations of Wolf Rayet stars, which yield (Abbott *et al.* 1986) nonthermal luminosities up to about $5 \; 10^{19}$ erg/sec/Hz at 5 GHz and thermal luminosities up to $7 \; 10^{18}$ erg/sec/Hz. Since the most important parameter, that enters here is the density of the wind, or in terms of wind parameters, the ratio of the mass loss to the wind speed $\dot{M}/V_W$, we induce that the most extreme stars have a higher value for $\dot{M}/V_W$ by 10.6. This translates into a higher nonthermal emission as well, by a factor of 3.3 to give $6.0 \; 10^{24}$ erg/sec/Hz $\eta_{-1} \, U_{1-2}^{2} \, B_{0.5}^{1.735}$, using the case when free-free absorption dominates (for synchrotron self absorption the corresponding luminosity is very nearly the same). This is very much more than the observed nonthermal luminosities and suggests that there is a limiting factor. We implicitly assume here, that some of the massive stars that exploded as the observed supernovae, either were Wolf Rayet stars before their explosion, or that their properties were not significantly different; since we derived the wind density from the timing of the lightcurve above, and the shock velocity both from the models of Owocki *et al.* (1988) and the argument that the shock velocity be larger than the Alfvén velocity, which in turn we argued is not very much lower than the wind velocity, the only other important parameter is the magnetic field strength, and that we assume then to be of similar magnitude. Since Wolf Rayet stars do not distinguish themselves from stars that somewhat later in life explode as supernovae, we can use all the same parameters, except for the shock speed. Then the only parameter left is the highly uncertain efficiency of electron injection $\eta$.

We can thus ask whether the efficiency $\eta$ might depend on shock speed. The shock speeds in supernovae are of order $0.03\,c$ or larger, while those in Wolf Rayet star winds are of order $0.01\,c$ or less. Somewhere between these speeds there appears to be a critical shock speed, at which the character of electron injection changes. Here we do not wish to discuss injection, but merely note that the data suggest efficiencies of order $\eta \simeq 10^{-6\pm 1}$ for the shock speeds thought to be normal in Wolf Rayet star winds, and of order $\eta \simeq 0.1$ for supernova explosions. The data thus suggest that the injection of electrons into the diffusive shock acceleration appears to exhibit a step function property at a critical shock velocity. Comparing normal supernova remnants, the nova GK Per, and the sources discussed above, a natural choice is a critical Alfvénic Machnumber, which we can only estimate to be in the range of 4 to 40. If true, this might be important also for electron injection in quasars, which also exhibit a dichotomy into radioloud and radioweak objects. In paper CR III, and below, we suggest that this concept leads to an estimate for the electron/proton ratio of the lower energy cosmic rays consistent with observations.

Now we can go back and ask again, whether free-free absorption or synchrotron self absorption dominates in the various cases considered: Clearly, when $\eta$, the efficiency for electron injection, is very small, then free-free absorption always dominates, and so the maximum luminosities during a radio variability episode of a Wolf Rayet (or OB) star should be frequency dependent. Also, for slow winds in supernova predecessor stars, again free-free absorption will dominate normally (*e.g.* in the two examples used above). On the other hand, for fast winds of the supernova predecessor stars (as in the case of a Wolf Rayet star exploding), synchrotron self absorption is likely to be stronger than free-free absorption and then the maximum luminosities at different radio frequencies should be independent of frequency. This is a testable prediction, since it relates radio and optical properties of a young supernova.

## D. OB stars

The theoretical luminosities derived can be compared with the observations of OB stars, which yield (Bieging *et al.* 1989) nonthermal luminosities up to about $7\ 10^{19}$ erg/sec/Hz at 5 GHz, and thermal luminosities up to $2.5\ 10^{19}$ erg/sec/Hz. Since the most important parameter, that enters here is the density of the wind, or in terms of wind parameters, the ratio of the mass loss to the wind speed $M/V_W$, we induce that the most extreme stars have a higher value for $M/V_W$ by 27.6. This translates into a higher nonthermal emission as well, by a factor of 5.4 to give $9.8 \, 10^{24} \, \text{erg/sec/Hz} \, \eta_{-1} \, U_{1,-2}^2 \, B_{0.5}^{1.735}$. This is very much more than the observed nonthermal luminosities. Since the shock properties are likely to be similar to Wolf Rayet stars, this is again consistent with the idea that the injection efficiency of electrons might be quite low, in conjunction with magnetic field values not too far from what we argued to be valid for Wolf Rayet stars.

We can also compare the detection statistics and observed spectral indices: Since in a variability episode a shock comes from below through the region of optical thickness near unity, the nonthermal emission increases rapidly to its maximum, when its spectral index is

$$\alpha(\nu, max) \; = \; -0.3 \, \alpha_{thin} \, , \qquad (133)$$

which is approximately $-0.2$; Thereafter, as the luminosity more slowly decreases, the spectral index gradually approaches the optically thin index of $-0.67$ or slightly steeper. Therefore we expect the spectral index distribution of detected sources to be a broad distribution from near $-0.2$ to near $-0.7$; this is what has been found (Bieging *et al.* 1989). Because of the two possible signs of the magnetic field orientation, and the associated drift energy gains for particles, we also expect that at any given time at most half of all sources are detectable with nonthermal emission, even at extreme sensitivity. This is consistent with the observations. We predict a similar behaviour for Wolf Rayet stars.

Here we have to ask how the magnetic fields can penetrate the radiative region from below. This question cannot presently be tackled by simulations on a very large computer yet, but it is likely that the circulations induced by rotation transport magnetic fields to the surface, where even a rather slight differential rotation draws out the magnetic field into a mostly tangential configuration. In this case, clearly the origin of the momentum of the wind can be readily accounted for from line driving (Lucy & Solomon 1970, Castor *et al.* 1975), and so we suspect that the magnetic driving adds only little (which is the "other" case discussed above near the end of the wind section).

We can also compare with the only existing theory to explain the nonthermal radio emission of single massive stars with winds, by White (1985): White's theory is based on a concept involving a large number of shocks and does not explain the data as already demonstrated by Bieging *et al.* (1989) in terms of i) radio spectral index, ii) time variability nor iii) of the statistics of detection. As a consequence any estimate of the strengths of magnetic fields based on White's theory is in doubt (see Bieging et al. 1989). The theory of diffusive particle acceleration by an ensemble of shock waves has been properly derived by Schneider (1993), and remains to be compared with the data of stars. Our theory is based on using single shocks and readily provides spectral indices in the entire range that is observed, it explains the time variability and easily accounts for a fair fraction of undetected sources.

Shocks in stellar winds as a consequence of supernova explosions give rise to $\gamma$-ray emission from hadronic interactions (Berezinsky & Ptuskin 1989); this latter conclusion was reached also for normal shocks in stellar winds by Chen & White (1991b) and White & Chen (1992). In the case of supernova explosions, it is important as a check, that the resulting predicted $\gamma$-ray production scales with the square of the wind-density, and thus with $(M/V_W)^2$ (Berezinskii *et al.* 1990, eq. 7.57 in VII§4).

The binary system WR140 has been detected by GRO (Hermsen *et al.* , seminar at the GRO symposium at Maryland, 1993; OSSE and BATSE instruments), confirming a detailed prediction by Eichler & Usov (1993), which is based on a model of colliding wind shocks. This latter agreement of observation and prediction is a consistency check on the concept, that freshly accelerated protons produce high energy photons from hadronic interaction, as opposed to purely leptonic processes.

## E. Radioemission from radiosupernovae

The adopted value for the magnetic field, however, was derived from the notion that the subsequent supernova shocks accelerate particles to extremely high energies. This argument can be checked with the observations of those supernovae of which the radioemission in the wind was really observed, five sources discussed by Weiler and colleagues in a number of papers (Weiler *et al.* 1986, 1989, 1990, 1991, Panagia *et al.* 1986) and the supernova 1987A (Turtle *et al.* 1987, Jauncey *et al.* 1988, Staveley-Smith *et al.* 1992) and related radio sources in starburst galaxies. However, we restrict ourselves to those supernovae for which we can reasonably assume that the predecessor star was indeed a star with a strong wind, and this we will do using statistical arguments in the subsequent section.

In fact, our model can be paraphrased as a numerical version of Chevaliers (1982) model with the parameters fixed: In the screen approximation valid for free-free absorption (see above for details), the time evolution of the nonthermal emission in Chevaliers model can be written as

$$L_\nu(nth) \ = \ K_1 \, \nu^\alpha \, t^\beta \, e^{-K_2 \, t^\delta} \tag{134}$$

with $\alpha$, $\beta$ and $\delta$ to be fitted to the data. For fast strong shocks our model predicts that $\alpha = \beta = -2/3$ and $\delta = -3$, and for slower shocks that still $\alpha = \beta$ but larger in number, for $U_1/V_W = 1$, $\alpha = \beta = -0.735$, for instance. This is consistent with the detailed fits for three supernovae (Weiler *et al.* 1986), which appear to arise from stars with stellar winds: Using the numbers from the fit using Chevalier's model, which requires $\beta = -3 + \alpha - \delta$, the averages are $\langle \alpha \rangle = -0.71 \pm 0.20$, $\langle \beta \rangle = -0.70 \pm 0.04$, $\langle \alpha, \beta \rangle = -0.70 \pm 0.13$, $\langle \alpha/\beta \rangle = 1.01 \pm 0.24$, and $\langle \delta \rangle = -3.01 \pm 0.17$ from these somewhat sparse data.

For SN 1986J the data clearly cover a sufficiently large time to test this numerical model in more detail; Weiler *et al.* (1989) find that this simple model in its screen approximation does not provide a good fit to the early epochs. We suspect similar to Weiler *et al.*, that this lack of a good fit can be traced to the simplification that we consider the external absorption as a simple screen, and disregard the lateral structure. Further possible reasons for a failure to strictly adhere to the simplified model are the following: a) The pre-shock wind may not be smooth in its radial behaviour, there might have been a weaker shock running through earlier which nevertheless can disturb the radial density profile (see the calculations by MacFarlane & Cassinelli 1989). b) Mixing between free-free absorption (outside the shock region) and synchrotron self-absorption (inside the shock region). c) The structure of acceleration in its latitude dependence is considered here only for the acceleration of particles (see paper CR I), but not for absorption and radiative transfer. Given a very detailed multifrequency data set it would be interesting to model the data fully.

In fact, we can use the numerical values for the radii for maximum luminosity derived above to obtain the pre-shock wind density, which is proportional to $\dot{M}/V_W$. With the data given by Weiler *et al.* (1986) this yields for the supernova 1979C

$$\frac{\dot{M}_{-5}}{V_{W,-2}}(1979\text{C}) \;=\; 3.0 \; 10^3 \, U_{1,-1.5}^{3/2},$$

and for the supernova 1980K

$$\frac{\dot{M}_{-5}}{V_{W,-2}}(1980\text{K}) \;=\; 3.4 \; 10^2 \, U_{1,-1.5}^{3/2}.$$

We note here that the nomenclature for these supernovae has changed between Weiler *et al.* 1986 and Weiler *et al.* 1989; we use here the more recent version.

Clearly, these numbers tell us that the predecessor stars had a slow wind, and thus were probably red supergiants (see Weiler *et al.* 1986). This then implies for shock speeds of $0.03\,c$, $\eta = 0.1$, and free-free absorption being dominant, that the nonthermal luminosities at 5 GHz expected versus observed are

$$L_{max}(1979\text{C}) \;=\; 3.0 \; 10^{27} \text{ versus } 2. \; 10^{27} \text{erg sec}^{-1}\,\text{Hz}^{-1},$$

and

$$L_{max}(1980K) = 9.0\,10^{26} \text{ versus } 1.\,10^{26} \text{erg sec}^{-1}\,\text{Hz}^{-1},$$

both for the assumed strength of the magnetic field. Since the luminosity is proportional to the magnetic field strength to the power 5/3, and directly proportional to the electron efficiency parameter $\eta$, we derive thus a lower limit to the magnetic field, given an upper limit on $\eta$. Since $\eta = 0.1$ is unlikely to be surpassed by much, using the ratio of the luminosities expected/observed of the two cases above yields an estimated lower limit to the magnetic field strength at our reference radius of

$$B > 1.5\,\eta_{-1}^{-3/5}\,U_{1,-1.5}^{-6/5}\,\text{Gauss.} \qquad (135)$$

This implies a strong lower limit to the magnetic field from using $\eta = 1$ of 0.4 Gauss. The uncertainty in these estimates is clearly at least a factor of 2.

The argument is often made, that hydrodynamic instabilities could increase the magnetic field strength in the postshock flow (Reynolds & Chevalier 1981); in such a case the estimate given here would only signify how much the magnetic field has increased behind the shock. However, the work by Galloway & Proctor (1992) suggests that dynamo time scales are not fast enough to do this. As emphasized in the summary, a direct observational check on the magnetic field strength in stellar winds is very desirable; it may be possible with data from a pulsar in a binary system with a massive early type star.

For 1987A, the initial radio luminosity was indeed of order $10^{25}$ erg/sec/Hz and so, applying our model with a shock speed of order 0.03 c is consistent with our expectation of then $3.4\,10^{25}\,\eta_{-1}$ erg/sec/Hz. Similarly, for the new radiosupernova 1993J in the galaxy M81 the observed maximum radioluminosity of $2.2\,10^{26}$ erg/sec/Hz at 99.4 GHz (Phillips & Kulkarni 1993) is quite compatible with a reasonable shock speed, and allowing for the fact, that the predecessor star was a red giant (thus $V_{W,-2} \ll 1$); both optically thin time evolution and optically thin spectrum are also consistent with the arguments presented here (Panagia *et al.* 1993).

This demonstrates, that strong magnetic fields also exist in the winds of massive stars in the red part of the Hertzsprung-Russell diagram, where the winds are slow - as argued earlier; note that the predecessor to supernova 1987A was a blue supergiant with a fast wind.

The only unknown parameter in all these predictions is the efficiency of electron acceleration $\eta$ and the strength of the magnetic field; with $\eta$ close to 0.1, clearly close to the maximum number reasonable, leads then to the requirement that the magnetic field strength is near to what we assumed, 3 Gauss at $10^{14}$ cm, but even higher, if $\eta$ is very much less than unity. This again confirms independently that indeed the magnetic field has to be as high as argued by Cassinelli (1982, 1991) to drive the winds and as is required to accelerate cosmic rays particles to energies near $3\,10^9$ GeV.

For supernova predecessor stars with fast winds like Wolf Rayet and OB stars, it is of interest to ask whether the expected luminosity violates the Compton limit, famous from the study of radioquasars. At the Compton limit the first order inverse Compton X-ray luminosity becomes equal to the synchrotron luminosity, and it has been found from observations that compact radioquasars are close to this limit and indeed have strong X-ray emission. This question can be formulated as a limit to the brightness temperature of the radio source (using here synchrotron self absorption) which then gives the limit

$$\eta_{-1} \left( \frac{\dot{M}_{-5}}{V_{W,-2}} \right) U_{1,-1.5}^2 \, B_{0.5}^{-1} \, \nu_{9.7}^{1.27} \; < \; 1.5\,10^7. \qquad (136)$$

For free-free absorption the limit is

$$\eta_{-1} \left( \frac{\dot{M}_{-5}}{V_{W,-2}} \right)^{-0.777} U_{1,-1.5}^2 \, B_{0.5}^{5/3} \, \nu_{9.7}^{-0.6} \; < \; 0.18. \qquad (137)$$

The first of these conditions is almost certainly always fulfilled, while the second one may be so tight as to suggest that inverse Compton X-rays might be observable. This has indeed been checked with modelling successfully the observed X-ray spectra beyond photon energies of 2 keV by Chen & White (1991a) for Orion OB stars. This suggests that the inverse Compton X-ray luminosity ought to be less, usually considerably less, than the Synchrotron luminosity. Most of the X-ray emission from hot stars is thought to arise from those same shocks in free-free X-ray emission which we consider for particle acceleration; a modelling of this was done by White & Long (1986) and MacFarlane & Cassinelli (1989).

## F. The statistics of Wolf Rayet stars and supernovae

There are a variety of ways to estimate the relative frequency of Wolf Rayet star supernova explosions relative to supernova explosions of lower mass stars (Hidayat 1991, Leitherer 1991, Massey & Armandroff 1991, Shara et al. 1991).

In our Galaxy there are between 300 and 1000 Wolf Rayet stars, which have an average lifetime of about $10^5$ years. This gives an estimated occurrence of Wolf Rayet supernovae of about one every 100 to 300 years. Since the total rate of supernovae in our Galaxy is estimated at about one every 30 years, this means that roughly one in 3 to one in 10 supernovae ought to represent the explosion of a Wolf Rayet star.

The numbers of stars on the main sequence between 8 solar masses and about 25 solar masses, and between 25 solar masses and the upper end of the main sequence also ought to correspond to the ratio of Wolf Rayet stars and the rest of those stars which explode as supernovae. Using for the simple estimate the Salpeter mass function gives here an estimated ratio of about 1 in 5 supernova events which originate from a Wolf Rayet star. 1 in 4 supernova events come from a star with a strong wind approximately (Wheeler 1989), where the change to stars with only a weak wind is estimated to be near a main sequence mass of 15 $M_\odot$.

The model for the origin of the high energy population of relativistic cosmic rays proposed in paper CR I and tested successfully in paper CR IV suggests from the energetics a ratio of about 1 in 3, assuming the amount of energy pumped into cosmic rays per supernova to be the same for all kinds. However, here we lump all supernova explosions into stellar winds together, both with fast and slow winds.

Hence in the sample of radio supernovae of type II (6 sources) presented by Weiler *et al.* (1986) there ought to be between none and two events based on Wolf Rayet star explosions; in the sample of radio sources likely also to be very young supernova remnants (28 sources with radio luminosities at 5 GHz) in the starburst galaxy M82 (Kronberg *et al.* 1985) there ought to be about at least between 3 and 10 sources which originate from a Wolf Rayet star explosion. On the other hand, all of these objects are possibly explosions of stars into former stellar winds (there is evidence that the initial mass function in starburst galaxies is biased in favor of massive stars), and so the proportion of the stars among them that are due to Wolf Rayet star explosions, could be even higher than estimated here. The radio luminosities are very similar for the sources in M82 and the radiosupernovae of Weiler *et al.* (1986); the average luminosity calculated in a variety of ways is always in the range of $3\ 10^{25}\,\mathrm{erg\,sec^{-1}\,Hz^{-1}}$ and $10^{26}\,\mathrm{erg\,sec^{-1}\,Hz^{-1}}$, fitting our expectation (see above) for $U_1 \simeq 0.03\,c$ rather well. We note, that for the same wind density, which is proportional to $\dot{M}/V_W$, the range of expected radio maximum luminosities is similar for Supernova explosions into slow and fast winds; on the other hand, if the mass loss rates are similar for slow and fast winds, then the densities can be much higher in slow winds, and the expected nonthermal radio luminosities are much higher for slow wind stellar explosions. Thus, observationally, we may only detect radiosupernovae of stars exploding into slow winds. There is one observational signature, which may be difficult to accurately measure: For a given mass loss rate, slow winds are usually dominated by free-free absorption, which leads to a frequency dependent maximum luminosity (see eq. 116), while fast wind are likely to be dominated by synchrotron-self-absorption, which leads to a frequency independent maximum radio luminosity. In Weiler *et* *al.* (1986) the data for the supernovae 1979C and 1980K illustrate this possible difference, where the best fit to 1979C shows a frequency dependent maximum luminosity, while for 1980K the maxima at 1.4 and 5 GHz appear to be the same.

For our arguments on cosmic rays it is not relevant whether the predecessor stars were Wolf Rayet stars or other massive stars with extended winds permeated by strong magnetic fields. We expect from the similarity in the internal structure of the stars on the upper main sequence, that their global properties such as the magnetic field strength generated should not be drastically different. Here we assume that we can use the implied properties for Wolf Rayet stars just as for stars of slightly lower mass which explode into slow winds.

## V. Electrons

### A. The injection of relativistic electrons

Above, we argued on the basis of a comparison of Wolf Rayet stars and radio supernovae, that there appears be a critical Alfvénic Machnumber for the injection of electrons. This critical Machnumber can not be pinpointed to one specific number at this time, but appears to be in the range of values of 4 to 40.

Levinson (1992, 1994) has argued on theoretical grounds, that indeed there is such a critical Alfvénic Machnumber for electron injection; his prediction contains a free parameter, so that only consistency with our argument can be verified.

Feldman (1993, priv.comm. at the Tucson meeting) noted, that data from solar wind shocks also show that there is critical Alfvénic Machnumber for electron injection (see Edmiston & Kennel 1984, Kennel *et al.* 1985 for a possible physical argument).

What we are lacking now, is a generalized theory to give this result, maybe even with a definite numerical value, as well as an argument, what determines or eliminates electron injection below this critical Alfvénic Machnumber. It is possible to interpret the radio data of stars with this concept, since it leads to a latitude restriction and thereby to a decrease in the radioemission; a quantitative check suggests that then the radio emission is confined to a fairly small polar region, while energetic protons can interact over a large part of the hemisphere (Nath & Biermann 1994b, Biermann, Strom, Falcke 1994).

### B. The maximum energy of relativistic electrons

For ISM-SN the maximum energy for electrons accelerated is given by about 30 - 100 GeV, using a low density interstellar medium, and Synchrotron losses versus net acceleration (acceleration minus adiabatic losses) as the limiting factor. It follows that the observed high energy electron tail cannot be attributed to normal supernova explosions into the interstellar medium.

For wind-SN the maximum energy is given by the consideration already used to derive the maximum emission frequency for nonthermal radioemission (above, eq. 122). The corresponding maximum electron energy is approximately given by

$$E_{e,max} \ = \ 0.3 \, r_{pc} \, B_{0.5}^{-2} \, U_{1,-2} \, \text{TeV}, \qquad (138)$$

where $r_{pc}$ is the radius where the stellar wind changes character, due to the shell formed by interaction with the interstellar medium. This radius may be of order a few parsec, and so the expected maximum energy, for a few parsec and a shock speed of order $10^4$ km/sec is about 3 TeV. This is consistent with the observations (summarized by Wiebel 1992) that detect energetic electrons up to a few TeV particle energies. We leave the problem of how to get such high energy electrons through the Galaxy to us and the implied consequences for another occasion.

We conclude for here, that for electrons above about 30 - 100 GeV, wind-supernovae are required to explain particle energies and spectrum.

## C. The spectrum of relativistic electrons

At particle energies of order GeV and slightly higher, the radioemission of external galaxies is the most reliable indicator of this spectrum; Golla (1989) has discussed the best data available, accounting for all the possible contribution from thermal radio emission, and finds, that all excellent data (*a sample of seven galaxies* with a well determined nonthermal radiospectrum with an error less than 0.1 in the spectral index) are compatible with a single spectral index of a powerlaw energy distribution for the electrons, in this energy range, of $2.76 \pm 0.12$. This is very nearly the same spectral index as found for Hydrogen in cosmic rays.

Wiebel (1992) compiled all the data available in the literature, and finds above about 30 GeV a spectrum of $E^{-3.26 \pm 0.06}$, which is to be compared with $E^{-3.33 - 0.02 \pm 0.02}$ as the expected spectrum from wind shocks of, steeper by unity than the injected spectrum due to synchrotron losses.

The particle energy of the switch between the two source sites, and the maximal electron energy observed, as well as the positron fraction, remain to be discussed in detail.

## D. The proton/electron ratio in cosmic rays

During the expansion the energy of any individual relativistic electron decreases by adiabatic expansion as the ratio of the radii from the time when electron injection ceases to the time of when proton injection ceases. After this point the energy densities of both particles decrease together. The simple dilution of the electron population is paralled by the steadily decreasing injection rate of protons, since both the volume and the ram pressure of the shock go down as radius to the -3 power during the adiabatic phase (see Shklovsky [1968], eq. 7.27); should the injection of protons cease at a later stage, then this differential dilution has also to be reckoned. The energy density ratio of the electron population relative to that of the protons is given by ($p = 2.420$)

$$(R_{crit,p}/R_{crit,e})^{p-1}. \tag{139}$$

This intermediate switch from electron injection with steady acceleration to a simple adiabatic loss regime determines the net scaling of the power of the electron population to that of the proton population in the cosmic rays. The observations suggest that from 1 GeV the energetic electron density is only about one percent (*e.g.* Wiebel 1992) of the density of the protons. From this observed ratio the energy density ratio integrated over the entire relativistic part of the particle spectrum, protons relative to electrons, for the spectral index of $-2.42$, is given by 4.3. In the standard leaky box model the ratio of energy densities of protons and electrons is not influenced by propagation effects. This suggests an expansion of a factor of order 3 in radius between the time when electron injection ceases and the time when proton injection ceases; here we assume that electrons and protons originally have comparable energy densities of their relativistic particle populations. We may have thus identified the origin of the observed electron/proton ratio in cosmic rays.

### E. The shell thickness of Supernova remnants in the ISM

One clear prediction of the concept of fast convective turbulence in the shock region is the fairly large thickness of the shell; this shell is the thickness of all the matter snowplowed together downstream, and in addition the length scale with the same column density upstream. Thus, what we refer to as *upstream* in our model is fully contained in the emission shell observed; the average shock location is deep inside the emission shell. The outer edge of the observed emission then, in this picture, is the location of the presently locally protuding shock, seen from the side. Therefore, the shell thickness in this model is both upstream and downstream in the language used earlier and, when referred to the outer radius, has the value

$$\Delta r/r = \frac{1 + U_1/U_2}{4 U_1/U_2} \tag{140}$$

for supernova explosions into the interstellar medium.

This gives $\Delta r/r = 5/16$ for a strong shock in a medium of constant density. It is important to note, that in the concept discussed here, this large thickness of the shell is traced to an instability of a cosmic ray mediated shock front, *i.e.* a shock strongly influenced in its structure by cosmic ray protons and other nuclei. As a consequence, a shock front, which does not inject new particles into the system, but just squeezes the existing energetic particle population, will not show this effect, and should therefore be much thinner, in the simple limit of a strong shock $\Delta r/r = 1/12$. This is indeed what is consistent with the Cygnus loop (Raymond, 1993, priv. comm.); however, one problem with the Cygnus loop is the open possibility that it is a warped sheet which may appear broader than it is really is.

The data for some sources can be read off published graphs (Pye *et al.* 1981, Dickel *et al.* 1982, Seward *et al.* 1983, Dickel *et al.* 1988, 1991), and are close to this value for parts of the sources Kepler, SN1006, Tycho, and RCW 103. Apparently, the new radio emission of supernova 1987A also appears to fit this prediction (Staveley-Smith *et al.* 1993). A more detailed analysis of such data is clearly required; it is obvious, that many supernova remnants are not nicely circularly symmetric, nor show a clear shell structure. After all, we know the interstellar medium to be extremely inhomogeneous. The data are also not always in agreement between radio and X-rays, which after all, trace nonthermal particles and thermal hot gas; when cooling becomes important, then the simple shell argument may not be sufficiently accurate, since it assumes constant density throughout the shell.

The analogous argument for winds remains to be done, and may be necessary to interpret the radio data for the nova GK Per.

## VI. Airshower data, other checks and consequences

We have discussed and reviewed the tests with airshower data elsewhere, and it suffices to summarize here the predictions and tests:

We predict for protons a spectrum of $E^{-2.75 \pm 0.04}$ (paper CR III); the Akeno data fit gives $E^{-2.75}$ (paper CR IV). Radiodata of normal galaxies give $E^{-2.76 \pm 0.12}$ (Golla 1989).

We predict for Helium and heavier elements $E^{-2.67 - 0.02 \pm 0.02}$ below the knee; the Akeno data also here give a spectrum very close to prediction, of $E^{-2.66}$.

We predict for the nuclei beyond the knee $E^{-3.07 - 0.07 \pm 0.07}$. The Akeno data give $E^{-3.07}$. The world data set of all good high energy data also gives this spectrum, as well as the cutoff at the predicted particle energy (paper UHE CR II). The Fly's Eye data (Bird *et al.* 1993) demonstrate that the chemical composition switches rapidly from a heavy composition to a light composition near $3\,10^{18}$ eV, as predicted already in 1990 (Biermann 1993d), and in a brief form earlier by Biermann & Strittmatter (1987).

We predict the particle energies at the bend, or knee, and the energies of the various cutoffs; the test with the Akeno data gives fitted values for these numbers close to prediction. From the fit the numerical values are rather strongly constrained, to within 20%, since we fit both vertical and slanted showers simultaneously in their showersize distribution.

It is obvious, that we have difficulty estimating the systematic error resulting from the quite general use of limiting arguments, such as always strong shocks, always maximal curvature in the elements of the fast convection, always ignoring possible enhancements of the magnetic field strength (important for the drifts). The predictions and fits, which appear to agree generally quite well, illustrate the possible refinements required within the context of the approximations made.

Further checks are possible with i) the data analysis of Seo *et al.* (1991), who give a spectrum of $E^{-2.74\pm0.02}$ for Hydrogen, and $E^{-2.68\pm0.03}$ for Helium, very close to our predictions; ii) The analysis of Freudenreich *et al.* (1990) who have argued for some time, that the chemical composition near the knee becomes heavily enriched; iii) the newest Fly's Eye data (Bird *et al.* 1994) which give a spectrum of $E^{-3.07\pm0.01}$ beyond the knee in the energy range $2\,10^{17}$ eV to $8\,10^{19}$ eV for the mono-ocular data, and $E^{-3.18\pm0.02}$ in the energy range $2\,10^{17}$ eV to $4\,10^{19}$ eV for stereo data, while the classical data from Haverah Park (Cunningham *et al.* 1980; also see Sun *et al.* 1993) give a spectrum of $E^{-3.09\pm0.02}$ below $10^{19}$ eV; and the iv) JACEE data, presented at Calgary (Asakimori *et al.* 1993), which also give spectra for Hydrogen and Helium at GeV particle energies, of $E^{-2.77\pm0.06}$ and $E^{-2.67\pm0.08}$, respectively.

Recently, we have discussed all the low energy data for these spectra (Biermann, Gaisser & Stanev 1994) and have shown, that also the overall normalization between ISM-SN and wind-SN is close to that expected on the basis of which stars have strong winds on the main sequence and which do not; this gives an overall ratio of energy contained in the two populations of 3 to 1. There we also attribute the underabundance of Hydrogen and Helium to a) the two different source sites, and b) the enrichment in the winds of evolved massive stars (see also Silberberg *et al.* 1990).

Also, in other recent work, we have shown that the low energy cosmic rays may reionize the intergalactic medium after leaving normal galaxies in galactic winds (Nath & Biermann 1993); and we have used the cosmic ray ionization rate implied by molecular cloud data to estimate the lower energy cutoff of the galactic cosmic rays to between 30 and 60 MeV kinetic energy (Nath & Biermann 1994).

Further more, we have used the concept of particle acceleration in shocks running through stellar winds and then hitting the surrounding molecular shells to propose an explanation for the strong $\gamma$-ray lines observed by COMPTEL in the Orion star forming region (Bloemen *et al.* 1994, Nath & Biermann 1994b; for competing models see Bykov & Bloemen 1994 and Ramaty *et al.* 1994). We consider this an important test of the entire picture, since we obtain a satisfactory match at once for the nonthermal radio emission of early type stars, the strong $\gamma$-ray line emission, and the very low $\gamma$-ray continuum emission. This successful match required a consideration of the latitude-dependent particle acceleration of a shock running through a stellar wind.

We conclude, that the model has passed a large number of tests quite successfully, allowing first quantitative checks to be made.

However, many questions remain to be answered, especially as regards the transport of cosmic rays, and the secondary to primary ratio.

## VII. Caveats

With our basic postulate of the *the smallest dominant scale* we have made a giant leap of faith in treating convection and turbulence in an ionized magnetic medium. This is the most glaring step in our argument, which it may take a long time to verify.

Similarly, in our derivation we used approximations from cosmic ray transport theory far beyond its proven range of validity. The formal agreement with the derivation of Drury (1983) gives reason to hope that we may not be too far from a proper description.

The agreement claimed between prediction and measurement of cosmic ray spectra critically depends on the correction for galactic transport, here derived from a Kolmogorov spectrum. The secondary to primary ratio in cosmic ray nuclei as a function of energy suggests clearly otherwise. We have not demonstrated that the structure of the interstellar medium and its temporal behaviour really allows the secondary to primary ratio to be understood in the context of our theory.

The data give conflicting evidence as to the chemical composition of the cosmic rays across the knee, whether it is dominantly light or increasingly heavy. We just show that the Akeno airshower data are consistent with the second possibility. However, it can be expected that MACRO-EASTOP data will shed considerable light on this issue.

For intergalactic transport of cosmic rays we have used an approximation of nearly straight line paths; a connection with the bubble structure of the galaxy distribution may exist. A comparison with the existing data base of high energy events is still outstanding.

For massive stars we have suggested a theory for the winds based on somewhat stronger magnetic fields than hitherto used, and have emphasized the possibility of a more tangential magnetic field geometry near the surface of the star. However, we have not actually calculated the structure of such a wind, nor have we been able to compare it with observations to the detail desirable.

While the apparent agreement of the predictions with data gives hope that it is worth pursueing the development of the theory, a lot of work remains to be done.

## VIII. Summary

Here we concentrate on the consequences for stars; the implications for cosmic rays have been described elsewhere (Biermann 1993b, c, d); the preceding section also gives the more important implications and further questions.

1) Novae, OB, and Wolf Rayet stars show evidence for particle acceleration in winds; the theory proposed can account for what is known of the spectra, luminosities, and temporal behaviour. Better and more complete data are needed.

2) White dwarfs provide a check on the notion, that a magnetic dynamo operates inside the inner convection zone in massive stars; this dynamo may provide the fairly strong magnetic field, which we argue is present in the stellar winds of massive stars.

3) Wolf Rayet stars may derive partially the momentum in their wind from the pressure gradient of the tangential magnetic field; their pressure gradient acts as an amplifier on a wind, which has some initial acceleration most likely from line-driving, possibly in a multiple scattering mode.

4) The comparison of the various stellar radio sources leads to the concept of a critical Alfvénic Machnumber for electron injection; this concept may be the basis for understanding of the proton/electron ratio in the observed galactic cosmic rays.

5) Supernova shocks in the stellar winds may provide the sources of galactic cosmic rays in a) Helium and heavier elements, b) electrons beyond about 30 GeV, c) all the way across the knee to about $3\,10^9$ GeV, where the chemical composition is mostly heavy. We have been able to quantitatively test the theory using airshower data and other recent cosmic ray data.

Various alternative models exist:

1. A postulated galactic wind model (Jokipii & Morfill 1987) may accelerate particles at a galactic wind termination shock; this is argued to contribute particles over the entire range of particle energies, from low energies to the end of the cosmic ray spectrum. At low energies, the distance which particles can reach upstream from the shock, when they stream back to our Galaxy, is limited by $\kappa/V_W$, where $V_W$ refers to the galactic wind velocity and $\kappa$ is the diffusion coefficient for energetic particles at the relevant energies. This distance is so small for any reasonable value for the diffusion coefficient, that low energy particles cannot reach the Galaxy. This requires that we have two different source populations, which merge near the knee, which in turn implies considerable finetuning of the source parameters. At the high particle energies, it is very difficult to see that the particles can be contained in the Galaxy (Berezinskii *et al.* 1990, IV§3).

2. The multiple shocks in the environment of OB superbubbles and young supernova remnants (Bykov & Toptygin 1990, 1992, Polcaro *et al.* 1991, 1993, Bykov & Fleishman 1992, Ip & Axford 1992) may also contribute.

3. The cosmic background of active galactic nuclei may contribute through the production of energetic neutrons which convert back to protons (Protheroe & Szabo 1992).

For all such models, a clear prediction of the spectrum and its chemical abundance distribution, as well as a detailed check with the airshower size distribution, both for slanted and oblique showers, is desirable.

We emphasize that our proposal, as far as the galactic cosmic rays are concerned, rests on a plausible but nevertheless speculative assumption about the nature of the transport of energetic particles in perpendicular shock waves, namely that there are large fast convective motions across the average location of the shock interface; this notion is, however, supported by radio polarization observations of young supernova remnants, as well as theoretical arguments, as briefly described above. The model has *predictive power*. We have given predictions and checks above. However, a large amount of work remains to be done.

There are many obvious tasks to be done next; as we have given a number of important steps and checks for the cosmic ray aspect elsewhere, we concentrate here on the ramifications concerning stars and stellar evolution:

1. Direct observational tests for the strength of the magnetic fields in OB stars, Wolf Rayet stars, red supergiant winds, and radio novae are critically important. The model presented above unequivocally depends on the magnetic field strengths proposed. One possibility to determine the strength of the magnetic field in winds appears to be the time dependent depolarization of the radio emission from a pulsar orbiting an upper main sequence star such as B1259-63 (Thorsett 1994 reporting on radio observations made by R.N. Manchester *et al.*).

2. The radio spectral evolution of novae, single OB and WR stars as well as radio supernovae, including the latitude distribution and radiative transfer, should be calculated to make the predictions more accurate, and thus more testable.

3. The dynamo producing the magnetic field in the inner convective zone in upper main sequence stars needs to be modelled as well as the transport of magnetic flux through the radiative zone outwards, in order to see whether the magnetic fields and their geometry proposed can really be generated and transported.

4. We need a detailed model of a stellar wind, which starts with, maybe, line driving in the multiple scattering mode, and then receives additional outward momentum from the gradient of the tangential magnetic field.

5. Finally, we need a proper stellar evolution calculation taking into account up to maximal rotation and up to maximum magnetic fields, which may play an important role in the final stages of stellar evolution.

## Acknowledgements

First of all, I wish to thank my graduate student H. Seemann for checking all equations in this article, and going over it with a finely toothed comb. Second, I have to thank my collaborators in the ongoing work partially reported here, Drs. J.P. Cassinelli, H. Falcke, T.K. Gaisser, J.R. Jokipii, H. Meyer, B.B. Nath, J. Rachen, E.-S. Seo, T. Stanev, R.G. Strom, and B. Wiebel. Further evolution of these ideas was prompted by intense discussions and exchanges again with my collaborators as well as Drs. G. Auriemma, V.S. Berezinsky, P. Bhattacharjee, J.H. Bieging, A. Bykov, P. Charbonneau, J. Cronin, L. Dedenko, J.R. Dickel, R. Diehl, V. Dogiel, J. Eilek, P. Evenson, M. Giller, F. Giovannelli, C. Jarlskog, J. Kirk, G. Mann, J.F. McKenzie, M. Nagano, S.P. Owocki, M.I. Pravdin, V.S. Ptuskin, R. Ratkiewicz, J. Raymond, E.R. Seaquist, F.D. Seward, V.V. Usov, G. Webb, J. Wdowczyk, and G. Zank. Helpful comments on various parts of this manuscript were received from V.S. Berezinsky, P. Charbonneau, J.R. Dickel, R.N. Manchester, J.F. McKenzie, S.P. Owocki, E.R. Seaquist, P. Song, J. Raymond, F.D. Seward, and G.P. Zank as well as two unknown referees. I noted the contribution from the interaction at various meetings above; I am grateful for the organizers for inviting me to their conferences, always in beautiful settings, such as the island Vulcano. Important help in finding some of the older references was received from Dr. J. Pfleiderer and the librarian of the MPI for Fluid Mechanics. High Energy Physics with the author is supported by a NATO travel grant (CRG 9100072).